\documentclass[aps,pre,citeautoscript,twocolumn,showpacs,floatfix,superscriptaddress,graphics,float]{revtex4-2}

\usepackage{amsmath,amssymb,graphicx,color,braket,hyphenat,makeidx,xcolor,dsfont,subfigure}
\usepackage[ocgcolorlinks,colorlinks=true,linkcolor=blue,citecolor=red,linktocpage=true]{hyperref}

\usepackage{natbib}

\newcommand{\tr}{\mathrm{Tr}}

\begin{document}

\title{Quantum coherence enables hybrid multitask and multisource \\ regimes in autonomous thermal machines}

\author{Kenza Hammam}

\affiliation{Centre for Quantum Materials and Technology, School of Mathematics and Physics, Queen’s University Belfast, Belfast BT7 1NN, United Kingdom}

\author{Gonzalo Manzano}
\affiliation{Institute for Cross-Disciplinary Physics and Complex Systems (IFISC) UIB-CSIC, Campus Universitat Illes Balears,E-07122 Palma de Mallorca, Spain}

\author{Gabriele De Chiara}
\affiliation{Centre for Quantum Materials and Technology, School of Mathematics and Physics, Queen’s University Belfast, Belfast BT7 1NN, United Kingdom}

\begin{abstract}
Non-equilibrium effects may have a profound impact on the performance of thermal devices performing thermodynamic tasks such as refrigeration or heat pumping. The possibility of enhancing the performance of thermodynamic operations by means of quantum coherence is of particular interest but requires an adequate characterization of heat and work at the quantum level. In this work, we demonstrate that the presence of even small amounts of coherence in the thermal reservoirs powering a three-terminal machine, enables the appearance of combined and hybrid modes of operation, where either different resources are combined to perform a single thermodynamic task, or more than one task is performed at the same time. We determine the performance of such coherence-enabled modes of operation obtaining their power and efficiency. In the case of hybrid regimes, the presence of coherence in the hot bath allows for an increase in power while maintaining high efficiencies. On the other hand, in combined regimes, a contrasting behavior emerges whereby coherence has a detrimental impact on power output and efficiency.
\end{abstract}

\maketitle

\section{Introduction}
Energy management and conversion entails leveraging temperature differences between reservoirs as a resource for generating heat flows that allow thermal machines to function, for instance, as a heat engine, as a heat pump or as a refrigerator~\cite{Kondepudi}. The development of efficient and adaptable thermal machines for controlling energy at the quantum scale poses considerable difficulties, but such an endeavour may be important for the advancement of novel quantum technologies in the future~\cite{AlexiaPRXQ2002}.

Quantum thermal machines \cite{KosloffARPC2014,benenti:2017,Levy_2018,MyersAVS2022,Arrachea2023} represent one of the focal points of fundamental research in the field of quantum thermodynamics \cite{goold:2016,Vinjanampathy_2016,QTM2018,Deffner_2019}, with extensive investigation conducted both theoretically~\cite{Scovil59,Alicki_1979,Palao01,linden:2010prl,Levy2012,brunner:2012,Correa14,Uzdin15,Campisi_2016,Silva16,Kilgour18,Manzano_2019,Saryal21,Diaz2021,daSilva22,Myersiop2022,saro_2023,aamir2023,guzman2023} and experimentally \cite{Brantut_2013,thierschmann:2015,Ro_nagel_2016,Maslennikov_2019,Peterson19,VonPRL_2019,Bouton_2021}.
These setups are usually characterised by a small quantum system acting as a working substance which is in contact with different thermal reservoirs and 
eventually influenced by external control. A significant hurdle lies in carefully determining how the quantum nature of these thermal devices affects the performance of thermodynamic tasks such as work extraction, heating and refrigeration.

In recent years, there has been a growing recognition of the significance of exploiting non-thermal features in non-equilibrium reservoirs for enhancing thermodynamic operation. In particular non-thermal reservoirs carrying quantum coherence \cite{Scully_2003,Mitchison_2015,Leggio15,Dag2016,Manzano19,rodrigues2019thermodynamics,Roman-Ancheyta_2020,Hammam21,Palafox2022}, squeezing \cite{Huang_2012,Rossnagel14,correa2014quantum,GonzaloPRE_2016,Niedenzu_2016,Agarwalla17,Manzano18b,Zhang23}, or classical and quantum correlations~\cite{Dillenschneider_2009,Park13,Francica2017,DeChiara20,Bresque21} have been proposed, leading to their experimental realizations~\cite{KlaersPRX2017,Zanin2022enhancedphotonic,Wang22,herrera2023correlationboosted}. Non-thermal features in the environment open up new possibilities for the optimisation of energy conversion in quantum systems and the enhancement of power and efficiency. Specifically, in the context of thermal machines, environmental quantum coherence can act as an extra source of work~\cite{GonzaloPRE_2016,Niedenzu_2018,rodrigues2019thermodynamics} or free-energy~\cite{Manzano19,Hammam21}, displaying unique characteristics as compared to standard (non-coherent) thermal reservoirs~\cite{Nonabelian22}. It can apparently boost heat engines efficiency above the standard thermodynamic Carnot bound~\cite{Scully_2003,Rossnagel14,Abah_2014}, causing the appearance of novel regimes of operation~\cite{GonzaloPRE_2016,Niedenzu_2016,hammam2022exploiting}. Moreover it has been found that coherence can also help in the thermalisation processes often appearing in cyclic engines~\cite{Dag2016,Roman-Ancheyta_2020,Tajima21}. However, the evaluation of the performance in many of these pioneering predictions relied on a different identification of the heat currents from the quantum reservoirs, which has been later revisited in different contexts~\cite{Bera17, Manzano18b, Niedenzu_2018,Manzano19,Ghosh2018,Niedenzu2019conceptsofworkin,rodrigues2019thermodynamics}.

Thermal machines models are often implemented in multi-terminal setups, 
that may lead to additional benefits, especially in the case of multiple conserved quantities. For instance, using a third terminal the separation of energy and charge flows becomes possible, thereby enhancing the efficiency of thermoelectric devices~\cite{serra2011,Jordan2012,Sanchez2015}. Three-terminal setups are also widely used platforms to amplify and modulate heat currents in quantum thermal transistors~\cite{JoulaiPRL2016,DuttaPRL2017,Gosh2022}. Another advantage of some multi-terminal devices is that they can perform more than a single thermodynamic task simultaneously. For instance, in Ref.~\cite{Entine2015} a three-terminal device was proposed where the invested thermal power from a bosonic bath might be employed to produce electric power and to refrigerate one electronic terminal. In Ref.~\cite{manzano2020hybrid} it has been shown that such \emph{hybrid regimes} can be implemented not only in state-of-the-art nanoscale thermoelectric energy harvesters~\cite{thierschmann:2015}, but are possible in generic autonomous thermal machines whenever multiple terminals and multiple globally conserved quantities (or charges) are combined. This observation has recently triggered the proposal of different devices leading to hybrid operational regimes~\cite{Lopez2023,Cavaliere2023,Jincheng2023}.

In this work, we explore the potential of non-thermal effects of the reservoirs as induced by a negligible amount of quantum coherence, in performing new regimes of operation in autonomous thermal machines, without the help of extra charges or external sources of work. To this aim, we study a prototypical model of a three-level autonomous machine in a three-terminal configuration~\cite{Scovil59,Palao01}. The three reservoirs are made of an ensemble of independent and identical units that interact sequentially with the three-level system representing the working machine. By injecting some amount of coherence in the units of one of the reservoirs, it becomes a source of both heat and work. As a result, new types of operating regimes emerge, namely, \textit{combined multisource regimes} where two input thermodynamic resources are employed to drive a useful task of the machine, and \textit{hybrid multitask regimes} where several thermodynamic tasks are performed simultaneously from a single input resource. Notably, these extra regimes would not be possible in the thermal case without further augmenting the charges in the setup or allowing external work to be performed. The performance of the machine is addressed through the characterisation of a recently introduced multi-task efficiency~\cite{manzano2020hybrid}, which helps us to accurately evaluate the beneficial or detrimental role of coherence in the performance associated with each regime. Our results indicate that while coherence might constitute an extra resource for allowing otherwise forbidden hybrid regimes, for most parameter choices its associated power-efficiency tradeoff leads to a lower performance as compared to the purely thermal case. Enhancements are possible only in some particular cases and are achieved in narrow parameter regions near the maximum power point.

Our paper is organised as follows. Section ~\ref{sec:secmodel} is dedicated to introducing the set up of our quantum thermal machine in a collisional framework. We derive its open quantum system dynamics in Secs.~\ref{sec:oqs}, and in Sec.~\ref{sec:thermoq}, we define all thermodynamic quantities  that will allow us to identify the regimes of operation at steady state in Sec.~\ref{sec:resultsregime}. Accordingly, we show the different regimes of operation when adding some amount of coherence in the state of the environment's units in Sec.~\ref{sec:regcoh}. This is then followed by the quantification of the performance of our three-terminal device by using the generic expression of the efficiency in Sec.~\ref{sec:performance} for the combined regimes (Sec.~\ref{sec:HP}) and the hybrid regimes (Sec.~\ref{sec:hybrid}). Finally, Sec.~\ref{sec:conclusion} is devoted to the conclusion.

\begin{figure*}[htp]
\begin{center}
\includegraphics[height=4.5cm]{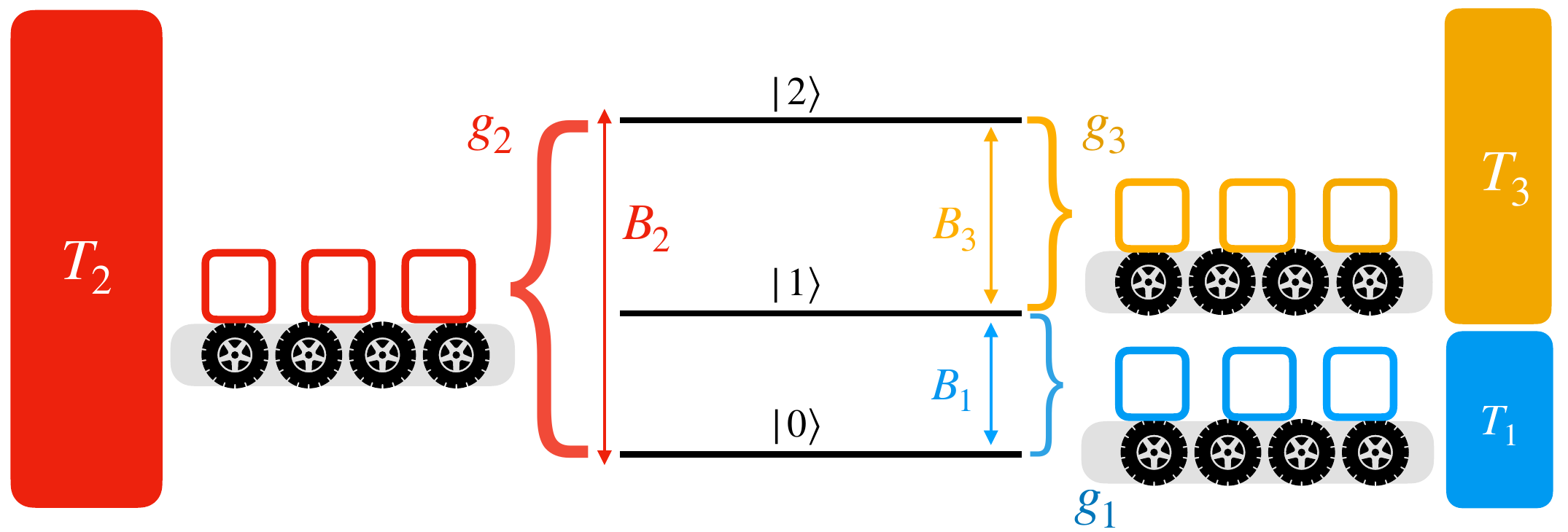}

\caption{Schematic model of a three-level autonomous thermal machine interacting in a three-terminal configuration. The three reservoirs are composed of a series of qubits which are initially identically prepared with some infinitesimal amount of coherence, see Eq.~\eqref{eq:rhoE}. They interact sequentially and resonantly with the transitions of the three-level system. }
\label{fig:coldbathcoherence}
\end{center}
\end{figure*}

\section{Thermal machine model}\label{sec:secmodel}
We consider a three-level system $S$ described by a three-dimensional Hilbert space spanned by the orthonormal basis $\{ |0\rangle, |1\rangle, |2\rangle\}$. It is in contact with an external environment comprising three different reservoirs whose interaction with the system is described within a collisional framework \cite{Campbell_2021,CICCARELLO20221,Cusumano2022}.
These reservoirs are represented by infinite series of auxiliary units---for simplicity we take qubits---which sequentially interact with the system one at a time. The interaction between the system and the reservoir units last for an interval $\tau$ (the collision time) and is associated to a succession of unitary operations, $U =e^{-i H \tau}$ ($\hbar=1$), where $H = H_S + \sum_{i=1}^3 (H_R^{(i)} + H_{S R} ^{(i)})$ is the total Hamiltonian comprising the system $H_S$, reservoir units $H_R^{(i)}$, and interaction $H_{S R}^{(i)}$ parts.

The reservoirs' units are described by the following local Hamiltonians:
\begin{equation}
    \quad H_{R}^{(i)}= \frac{B_i}{2} \sigma_{z}^{(i)}, ~~ i= 1, 2, 3,
\end{equation}

where $\sigma_{x,y,z}^{(i)}$ represent the Pauli matrices for the environmental auxiliary units and $B_i$ is the corresponding applied magnetic field, which we assume to be resonant with the three-level system transitions, such that
\begin{equation}
H_S= B_{2} \ket{2}\bra{2} + B_{1} \ket{1}\bra{1},
\end{equation}
with the consistency condition $B_3 = B_2 - B_1$.

We assume the system to interact with two-level units coming from cold, intermediate and hot reservoirs at different temperatures $T_1$, $T_2$, and $T_3$, respectively, with $T_1 < T_2 < T_3$. Figure ~\ref{fig:coldbathcoherence} illustrates the transitions in the system coupled with each reservoir: $(1)$ the transition $|1\rangle \leftrightarrow |0\rangle$ with the cold bath at $T_1$ units; $(2)$ the transition $|2\rangle \leftrightarrow |0\rangle$ with the intermediate temperature bath at $T_2$ and $(3)$ the transition $|2\rangle \leftrightarrow |1\rangle$ with the hot bath at $T_3$.

The interaction between the system and the auxiliary units of the baths is assumed to be weak, and the corresponding interaction Hamiltonians are of the general form:

\begin{align} 
    H_{S R}^{(1)}&= \frac{g_1}{\sqrt{\tau}}(\ket{0}\bra{1} \sigma_1^{\dagger} + \ket{1}\bra{0} \sigma_1) \nonumber \\
     H_{S R}^{(2)}&= \frac{g_2}{\sqrt{\tau}}(\ket{0}\bra{2}\sigma_2^{\dagger} + \ket{2}\bra{0} \sigma_2) \nonumber \\
      H_{S R}^{(3)} &= \frac{g_3}{\sqrt{\tau}}(\ket{1}\bra{2}\sigma_3^{\dagger} + \ket{2}\bra{1} \sigma_3)
\end{align}
where $\sigma_i=\frac{1}{2}(\sigma_{x}^{(i)}-i \sigma_{y}^{(i)})$, and $g_i$ denotes the coupling strength.

 The auxiliary units of the reservoirs are all initialized in the following fixed state bearing coherences in the $H_R^{(i)}$ eigenbasis~\cite{rodrigues2019thermodynamics}:
\begin{equation} \label{eq:rhoE}
    \rho_{R}^{(i)}= \frac{e^{-\beta_i H_R^{(i)}}}{Z_i} + \lambda_i\sqrt{\tau} \chi_i,
\end{equation}
where the first term corresponds to the thermal Gibbs state at (inverse) temperature $\beta_i =1/T_i$ (we set Boltzmann's constant as $k_B=1$), such that $Z_i=\mathrm{Tr}[e^{-\beta_i H_R^{(i)}}]$ is the partition function, and in the second term $\chi_i =\cos{\phi_i}\sigma_x^{(i)}+\sin{\phi_i}\sigma_y^{(i)}$ is a trace-less Hermitian operator with zero diagonal elements in the energy basis of $H_R^{(i)}$. Here the angle $\phi_i$ is the azimuth of the environmental unit’s Bloch vector. 

Notice that the state above deviates from a thermal Gibbs state by the inclusion of coherences whenever $\lambda_i \neq 0$ by an amount proportional to the square root of the collision time $\tau$. We note that we are interested in small amounts of coherence, which refers to the situation where  $\tau$ approaches zero. In this case, the magnitude of the second term of Eq.~\eqref{eq:rhoE} is significantly smaller compared to the first term and any choice of $\chi_i$ with zero diagonal elements is permitted. On the other hand if $\tau$ is finite, a positive semidefinite $\rho_{R}^{(i)}$ is not guaranteed considering any choices of $\chi_i$.

\subsection{The open quantum system dynamics}\label{sec:oqs}
As it is costumary in collisional models \cite{CICCARELLO20221}, we assume no initial correlations between the system and the units of the reservoirs at the beginning of each interaction, which means that they start in a product state $\rho_S(t)  \bigotimes_{i=1}^{3} \rho_{R}^{(i)}$ such that $\rho_S(t)$ is the state of the system at a generic time $t$ and $\rho_{R}^{(i)}$ are fixed. Therefore, for a given collision between the system and the environment, the global state of system and reservoirs changes to:
\begin{align} \label{eq:rhoSR}
    \rho_{S R}(t+\tau)= U~\left( \rho_S(t)  \bigotimes_{i=1}^{3} \rho_{R}^{(i)} \right)~ U^{\dagger}.
\end{align}
A Baker-Campbell-Haussdorf series expansion of Eq.~\eqref{eq:rhoSR} in $\tau$ is applied
\begin{widetext}
\begin{align}
&\rho_{S R}(t+\tau) = \rho_S(t)  \bigotimes_{i=1}^{3} \rho_{R}^{(i)}-i \tau \left[H,\rho_S(t)  \bigotimes_{i=1}^{3} \rho_{R}^{(i)} \right] 
-\frac{\tau^2}{2}\left[H,\left[H,\rho_S(t)  \bigotimes_{i=1}^{3} \rho_{R}^{(i)}\right]\right] 
\nonumber \\
&=\rho_S(t)  \bigotimes_{i=1}^{3} \rho_{R}^{(i)} 
-i \tau \left[H_S+\sum_{i=1}^{3}H_{R}^{(i)},\rho_S(t)  \bigotimes_{i=1}^{3} \rho_{R}^{(i)} \right] \\ \nonumber
&-i\tau \left[\sum_{i=1}^{3}H_{SR}^{(i)},\rho_S(t)  \bigotimes_{i=1}^{3} \rho_{R}^{(i)} \right] 
- \sum_{i=1}^{3}\frac{\tau^2}{2}\left[H_{SR}^{(i)},\left[H_{SR}^{(i)},\rho_S(t)  \bigotimes_{i=1}^{3} \rho_{R}^{(i)}\right]\right]
\end{align}
\end{widetext}
which, after tracing out the reservoir states, leads to a discrete map over the reduced system's state while keeping only terms which are linear in $\tau$:
\begin{align}
    &\rho_S(t+\tau) = \tr_R\left[ \rho_{S R}(t+\tau)\right] =\rho_S(t) - i\tau \left[H_S,\rho_S(t) \right]\nonumber \\
    &- i\tau \left[\sum_{i=1}{G}_S^{(i)},\rho_S(t) \right] +\tau \sum_{i=1}\mathcal{D}_i[\rho_S(t)] + O(\tau^2),
\end{align}
where we introduced the system operators $G_S^{(i)} := \tr_{R}\left[H_{S R}^{(i)}  \rho_{R}^{(i)}\right]$ and the Lindblad superoperators $ \mathcal{D}_i(\rho_S) :=  - \frac{\tau}{2} \tr_{R} \left[H_{SR}^{(i)}, [H_{S R}^{(i)}, \rho_S \rho_{R}^{(i)}] \right].$

In the continuous limit, $\tau \to 0$, the changes in the system density operator $\Delta \rho_S := \rho_S(t+\tau) - \rho_S(t)$ become smooth by dividing each side with $\tau$:
\begin{equation}
    \dot\rho_S(t)= \lim_{\tau \rightarrow 0}\frac{\Delta \rho_S}{\tau},
\end{equation}
leading to the following Markovian master equation \cite{rodrigues2019thermodynamics}:
\begin{equation}
\dot\rho_S(t) = -i[H_S+\sum_{i=1}{G}_S^{(i)},\rho_S(t)]+\sum_{i=1}^{3} \mathcal{D}_i[\rho_S(t)]
\end{equation}
where we obtain three effective Hamiltonian terms induced by the interaction with the reservoirs, given by the operators:

\begin{align} 
G_S^{(1)} &= \lambda_1 g_1 (e^{i\phi_1}\ket{0}\bra{1}+ e^{-i\phi_1}\ket{1}\bra{0}), \nonumber\\ 
G_S^{(2)} &= \lambda_2 g_2 (e^{i\phi_2}\ket{0}\bra{2}+ e^{-i\phi_2}\ket{2}\bra{0}), \\ \nonumber
G_S^{(3)} &= \lambda_3 g_3 (e^{i\phi_3}\ket{1}\bra{2}+ e^{-i\phi_2}\ket{2}\bra{1}). \nonumber
\end{align}
These extra terms provide a coherent contribution to the three-level system dynamics which is present only whenever the reservoir contains coherence, $\lambda_i \neq 0$. 

Moreover, we obtain the three following Lindbladian dissipative terms corresponding to each reservoir incoherent contribution to the dynamics:
\begin{align}
\mathcal{D}_1(\rho_S) &=  \gamma^{-}_{1} (\ket{0}\bra{1}\rho_S \ket{1}\bra{0}-\frac{1}{2}\{\ket{1}\bra{1}, \rho_S\}) \\ \nonumber &~+ \gamma^{+}_{1} (\ket{1}\bra{0} \rho_S \ket{0}\bra{1} -\frac{1}{2}\{\ket{0}\bra{0}, \rho_S\}), \nonumber \\ 
\mathcal{D}_2(\rho_S) &=   \gamma^{-}_{2} (\ket{0}\bra{2} \rho_S \ket{2}\bra{0}-\frac{1}{2}\{\ket{2}\bra{2}, \rho_S\}) \\ \nonumber &+ \gamma^{+}_{2} (\ket{2}\bra{0} \rho_S \ket{0}\bra{2} -\frac{1}{2}\{\ket{0}\bra{0}, \rho_S\}), \nonumber \\ 
\mathcal{D}_3(\rho_S) &= \gamma^{-}_{3} (\ket{1}\bra{2} \rho_S \ket{2}\bra{1}-\frac{1}{2}\{\ket{2}\bra{2}, \rho_S\})\\ \nonumber &+ \gamma^{+}_{3} (\ket{2}\bra{1} \rho_S \ket{1}\bra{2} -\frac{1}{2}\{\ket{1}\bra{1}, \rho_S\}), 
\end{align}
with $\gamma^{+}_{i}= |g_i|^2 \langle \sigma_i^{\dagger} \sigma_i\rangle$ and $\gamma^{-}_{i}= |g_i|^2 \langle \sigma_i \sigma_i^{\dagger} \rangle$ for $i=1,2,3$ describing the rates at which incoherent jumps among the three energy levels of the machine are triggered by the reservoirs. Using the expression of the initial state of the reservoir units, Eq.~\eqref{eq:rhoE}, we have $\langle \sigma_i^\dagger \sigma_i \rangle = 2\bar{n}_i + 1$, where $\bar{n}_i := [\exp{(\beta_i B_i)}-1]^{-1}$. These rates verify local detailed balance $\gamma^{+}_{i} = \gamma^{-}_{i} e^{-\beta_i B_i}$ and can be rewritten in standard form, $\gamma^{+}_{i}= \gamma_i \bar{n}_i$ and $\gamma^{-}_{i}= \gamma_i (\bar{n}_i + 1)$ by setting $\gamma_i = |g_i|^2/(2 \bar{n}_i + 1)$.

Analogously, the evolution of the $i$th reservoir units as they interact with the three-level machine can be obtained from Eq.~\eqref{eq:rhoSR} by tracing out the degrees of freedom of the machine and of all the reservoirs except the $i$th one, which we label as ${\rm Tr}_{\bar{i}}[...]$. The state of the reservoirs' units after each collision, keeping terms up to second order in $\tau$, reads:
\begin{align}\label{eq:rhoEprime}
\rho^{\prime^{(i)}}_R &= \tr_{\bar{i}}\left[\rho_{SR}(t + \tau) \right] = \nonumber 
\\
&=\rho_R^{(i)} -i \tau \left[G_{R}^{(i)}, \rho_R^{(i)}\right]  +  \tau~\mathcal{D}_R(\rho_R^{(i)}) + O(\tau^2),
\end{align}
where we obtain an effective Hamiltonian correction for each bath $G_{R}^{(i)}:=\tr_{\bar{i}}\left[H_{S R}^{(i)} \rho_{S}\right]$ given by the following operators: 
\begin{align} 
G_R^{(1)} &= g_1 (\rho_{1,2}\ket{0}\bra{1}+ \rho_{2,1}\ket{1}\bra{0}) \nonumber\\ 
G_R^{(2)} &=  g_2 (\rho_{1,3}\ket{0}\bra{2}+ \rho_{3,1}\ket{2}\bra{0}) \\ \nonumber
G_R^{(3)} &=  g_3 (\rho_{2,3}\ket{1}\bra{2}+ \rho_{3,2}\ket{2}\bra{1}), \nonumber
\end{align}
with $\rho_{i,j}= \bra{i} \rho_S(t) \ket{j}$ the entries of the three-level system's density matrix in the basis $\{ |0\rangle, |1\rangle, |2\rangle\}$, and the corresponding dissipative contribution $\mathcal{D}_i^R(\rho_R^{(i)}) :=  - \frac{\tau}{2} \tr_{\bar{i}}\left[H_{SR}^{(i)},[H_{S R}^{(i)}, \rho_S \bigotimes_{j=1}^{3} \rho_{R}^{(j)} \right]$.

The collision of the three-level machine with the environmental units results, in the long time limit, in driving the machine to a nonequilibrium steady state (NESS) $\rho_S^{NESS}$ (see Appendix \ref{steadystate}). The presence of a coherent exchange of energy between the three-level system and the reservoirs leads to the machine acquiring coherence in the $H_S$ basis in such limit despite the decoherence induced by thermal contacts. This implies that $\rho_S^{NESS}$ depends explicitly on $\lambda_i$. Moreover, we remark that whenever the NESS is reached, all forthcoming auxiliary units from the reservoirs undergo the same identical evolution.

 \subsection{The thermodynamic quantities}\label{sec:thermoq}

From the evolution of the machine and the reservoir units obtained above, we are now in a position to evaluate all relevant thermodynamic quantities of the setup. In order to verify the autonomous character of the machine model, we first evaluate the total energy change during a collision:
\begin{equation}
\label{eq:Wdot}
W_\mathrm{mec}= \tr\left[\left(H_{S}+H_R\right) \Delta \rho_{SR}(t) \right] = 0,
\end{equation}
with $\Delta\rho_{SR}(t) =\rho_{SR}(t + \tau) -\rho_S(t)\bigotimes_{i=1}^{3}\rho_R^{(i)}$ the change in the global state. The quantity $W_\mathrm{mec}$ plays the role of the mechanical work cost of switching on and off the interaction between system and reservoirs~\cite{Barra15,Strasberg17,DeChiara_2018}.
Here we obtain $W_\mathrm{mec} = 0$ as follows from the strict conservation of energy during the global unitary evolution describing the collisions, $[U, H_S + H_{R}] = 0$, as enforced by the resonance condition for the energy spacings of the three-level system and the auxiliary units.

The energy changes in the machine and reservoir $i$ due to a single collision are given by: 
\begin{align}
\label{eq:Qdot}
\Delta E_S  &:= \tr\left[H_{S} ~\Delta \rho_{SR}(t) \right] , \\ \label{eq:Edot}
\Delta E_R^{(i)} &:=\tr[H_{R}^{(i)} ~\Delta \rho_{SR}(t)],     
\end{align}
so that energy conservation is satisfied, $\Delta E_S + \sum_{i=1}^{3} \Delta E_R^{(i)} =  W_\mathrm{mec} = 0$. It is worth remarking that, since the initial states of the units in the reservoirs are not, in general, thermal states, we should refrain from the standard identification of all the energy exchanged with them as heat~\cite{esposito2010lindenberg}, but in the following we will split this quantity into work and heat contributions, acknowledging the nonequilibrium character of the reservoirs.

The second law of thermodynamics in the setup can be expressed by the non-negativity of the average entropy production $S_\mathrm{tot}= \mathcal{I}^{\prime}(S:R)+ \sum_{i=1}^3 S(\rho_{R}^{\prime~ (i)}||\rho_{R}^{(i)})  \geq 0$ due to collisions with the three reservoirs.
The mutual information between the system and the auxiliary units after the collision is given by $\mathcal{I}^{\prime}(S:R)=S(\rho_S^{\prime})+S(\rho_R^{\prime})-S(\rho_{SR}^{\prime})$. Since $S$ and $R$ are initially uncorrelated ($\mathcal{I}(S:R)=0$), we obtain:
\begin{equation}
    \mathcal{I}^{\prime}(S:R)=\Delta S_\mathrm{sys} + \sum_{i=1}^3 \Delta S_{R}^{(i)},
\end{equation}
where $\Delta S_\mathrm{sys} = S[\rho_{S}(t+\tau)]-S[\rho_{S}(t)]$ and $\Delta S_{R}^{(i)}=S(\rho_{R}^{\prime~ (i)})-S(\rho_{R}^{(i)})$ are the entropy changes in the three-level system and the auxiliary units from reservoir $i$, respectively. On the other hand, the terms $S(\rho_{R}^{\prime~ (i)}||\rho_{R}^{(i)})$ stand for the displacement of the reservoir units from their initial states during each single collision, which are ultimately ``reset" back to their original states. Above we denoted by $S(\rho) = -\tr\rho\ln\rho$ the von Neumann entropy of state $\rho$ and $S(\rho || \sigma)=\tr \rho(\ln\rho-\ln\sigma)$ the relative entropy between states $\rho$ and $\sigma$, which is positive and becomes zero only if $\rho = \sigma$.

The expression of the entropy production, for a single dynamical step $\tau$, is~\cite{esposito2010lindenberg,Manzano18,landi2021irreversible}:
\begin{align}\label{eq:entro}
    S_\mathrm{tot}=\Delta S_\mathrm{sys} + \sum_{i=1}^3 \left( \Delta S_{R}^{(i)} + S(\rho_{R}^{\prime~ (i)}||\rho_{R}^{(i)}) \right).
\end{align}

The entropy production in Eq.~\eqref{eq:entro} can be written in standard thermodynamic form as the change in the system entropy and the sum of the entropy exchanged with the reservoirs:
\begin{equation}\label{eq:entro2}
    S_\mathrm{tot} = \Delta S_{\mathrm{sys}} - \sum_i \beta_i Q_{i},
\end{equation}
by identifying the heat exchanged with the collisional reservoirs as:

\begin{align} \label{eq:heat}
    Q_{i} &:= - T_i [\Delta S_{R}^{(i)} + S(\rho_{R}^{\prime~ (i)}||\rho_{R}^{(i)})] \nonumber \\ 
    &~=  T_i \tr[\Delta \rho_R^{(i)} \ln \rho_R^{(i)}], 
\end{align}
with $\Delta \rho_R^{(i)} = \rho_R^{\prime~ (i)}-\rho_R^{(i)}$, in analogy to Refs.~\cite{Manzano18b,rodrigues2019thermodynamics}. This identification of heat, instead of the expression $Q_{i} =- T_i \Delta S_{R}^{(i)}$, as often employed in similar situations~\cite{Bera17,Hammam21}, responds to the fact that here we are implicitly assuming that the reservoir units become inaccessible after colliding with the machine, hence leading to an effective resetting to their initial states~\cite{Manzano18,landi2021irreversible}. 

\begin{table*}[t]
\centering
\begin{tabular}{|c|c|l|}
\hline
I & Quantum absorption refrigerator & $\Dot{W} = 0, \dot Q_1>0,\dot Q_2<0, \dot Q_3>0$ 
\\
\hline
II & Heat pump & $\Dot{W} = 0, \dot Q_1<0,\dot Q_2>0, \dot Q_3<0$ 
\\
\hline
III & Power and heat driven refrigerator & $\Dot{W} > 0, \dot Q_1>0,\dot Q_2<0, \dot Q_3>0$ 
\\
\hline
IV & Power and heat driven pump & $\Dot{W} > 0, \dot Q_1<0,\dot Q_2>0, \dot Q_3<0$ 
\\
\hline
V & Hybrid power driven refrigerator and heat pump & $\Dot{W} > 0, \dot Q_1>0,\dot Q_2<0, \dot Q_3<0$ 
\\
\hline
VI & Hybrid heat engine and heat pump & $\Dot{W} < 0, \dot Q_1<0,\dot Q_2>0, \dot Q_3<0$ 
\\
\hline
VII & Dual sink accelerator & $\Dot{W} > 0, \dot Q_1<0,\dot Q_2<0, \dot Q_3>0$ 
\\
\hline
VIII & Triple power driven pump & $\Dot{W} > 0, \dot Q_1<0,\dot Q_2<0, \dot Q_3<0$ 
\\
\hline
\end{tabular}
\caption{\small Operational regimes for a quantum thermal machine operating with three collisional baths. The first column represents the label corresponding to each regime of operation of the machine which is given in the second column. The last column further characterises the regimes by showing the signs of the coherent work and the heat flows to each reservoir.} \label{table:functionings}
\end{table*}

Inserting the explicit expression for $\Delta \rho_R^{(i)}$ obtained from Eq.~\eqref{eq:rhoEprime} into Eq.~\eqref{eq:heat} we obtain:
\begin{align}\label{eq:27}
    {Q}_i  =  - \Delta E_R^{(i)} - W_i.
\end{align}
The second term in Eq.~\eqref{eq:27} is a work contribution from reservoir $i$ due to the presence of coherence in its initial state:
\begin{equation} \label{eq:coherentwork}
{W}_i := -i\lambda_i \beta_i \tau \langle [G_R^{(i)},H_{R}^{(i)}]\rangle_{\chi_i},    
\end{equation}
which arises from the unitary part of the evolution in Eq.~\eqref{eq:rhoEprime}. Above we denoted by $\langle ... \rangle_{\chi_i}=\tr \{(...) \chi_{i}\}$ the trace over the coherent part of the initial state of the auxiliary units.

The above identifications leads to the formulation of the first law of thermodynamics in the setup:
\begin{equation}\label{eq:firstlaw}
    \Delta E_S = \sum_i (Q_i + W_i).
\end{equation}
It is worth remarking that not all the energy exchanged with the environmental units can be considered as heat, but we obtain a second term in Eq.~\eqref{eq:27} of genuine quantum coherent origin, which would disappear for a thermal environment $(\lambda_i = 0)$. In other words, $Q_i$ above is the part of the energy exchanged with the environment $i$ during collisions that, according to Eq.~\eqref{eq:entro2}, is connected to entropy exchanges with that reservoir, while $W_i$ can be identified with the energy input from the reservoir not contributing to any entropy flux from or to the environment. When the NESS is reached, the energy of the three-level system does not change anymore, $\Delta E_S = 0$, and the first law above becomes $\sum_i (Q_i + W_i)= 0$.

Moreover, as noticed in Ref.~\cite{rodrigues2019thermodynamics}, when $\tau \ll 1$, the expression for the relative entropy between the final and initial states of the auxiliary units of the reservoir appearing in Eq.~\eqref{eq:entro} becomes:
\begin{equation}\label{eq:coh}
S(\rho^{\prime~ (i)}_R||\rho_R^{(i)})= \Delta C_R^{(i)} + {W}_i \geq 0,
\end{equation}
where $\Delta C_{i}= C(\rho^{\prime~ (i)}_{R})- C(\rho_R^{(i)})$ is the change in relative entropy of coherence~\cite{Baumgratz2014,Streltsov2017} in the state of the auxiliary units, $C(\rho) := S(\bar{\rho})-S(\rho)$ with $\bar{\rho}$ is the diagonal part of state $\rho$ in the $H_R^{(i)}$ basis. This expression helps us to understand the relation of the coherent work contribution from the reservoirs to the actual coherence present in the auxiliary units. Indeed by virtue of the inequality in Eq.~\eqref{eq:coh}, a decrease in the coherence of the reservoir units $\Delta C_R^{(i)} \leq 0$ gets linked to the performance of work by the reservoir, ${W}_i \geq 0$, while work extraction, ${W}_i \leq 0$, implies the generation of coherence in the reservoir units $\Delta C_R^{(i)} \geq 0$. See also Ref.~\cite{Manzano19} about the generation of coherence by thermal resources.

Taking the limit $\tau \rightarrow 0$, we can immediately obtain the corresponding expressions for the rate of energy change in the system, $\dot E_S(t):=\lim_{\tau\to 0} \Delta E_S/\tau$, heat currents to reservoir $i$, $\dot{Q}_i = \lim_{\tau\to 0} Q_i/\tau$ and coherent power from reservoir $i$, $\dot{W}_i := \lim_{\tau\to 0} W_i/\tau$. We report the detailed analytical expressions of these quantities in Appendix~\ref{ap:power}.

When the NESS is reached the second law inequality in the setup is obtained from the non-negativity of the entropy production rate $\dot{S}_\mathrm{tot} :=\lim_{\tau\to 0} S_\mathrm{tot}/\tau \geq 0$, which imposes
\begin{equation}
    \sum_i \beta_i \dot{Q}_i \leq 0.
\end{equation}
The above second-law inequality can be rewritten in an appealing way for the cases in which only one of the baths in the setup contains coherence (e.g. the $k$-th bath):
\begin{equation}
    \dot{W}_k + \sum_{i \neq k} (\beta_k - \beta_i) \dot{Q}_i \geq 0.
\end{equation}
This inequality suggests that when the presence of coherence in one of the reservoirs leads to a positive coherent power $\dot{W}_k > 0$, it can be used to power heat currents to (from) other reservoirs, $\dot{Q}_i<0$ ($\dot{Q}_i>0$), even against a temperature gradient, i.e. when $\beta_k > \beta_i$ ($\beta_i > \beta_k$). For example, coherent power $\dot{W}_k$ may allow refrigeration of a cold bath with $\beta_i > \beta_k$, extracting a heat current $\dot{Q}_i > 0$ from it, or allow heat pumping into a hot bath $\beta_i < \beta_k$ with $\dot{Q}_i < 0$. Remarkably the presence of coherence allows also the appearance of hybrid regimes where more than one useful task can be performed simultaneously, or combined regimes where different resources can be combined to perform a single task. A detailed account of the possible regimes of operation of the device under the presence of coherence is given in next section. 

In the linear response regime, we can see that the effect of coherence within the environmental units is quadratic, such that both incoherent heat currents and coherent work contributions can be rewritten in the form of a Maclaurin series expansion up to the first order in $\lambda_1^2$:
\begin{align} 
    \Dot{Q}_i &\approx  \Dot{Q}_i|_{\lambda_i=0} +\frac{\lambda_i^2}{2!}\frac{\partial^2 \Dot{Q}_i}{\partial^2 (\lambda_i^2)}|_{\lambda_i=0}, \label{eq:dotqiseries} \\
    \Dot{W}_i &\approx  \Dot{W}_i|_{\lambda_i=0} + \frac{\lambda_1^2}{2!}\frac{\partial^2 \Dot{W}_i}{\partial^2 (\lambda_i^2)}|_{\lambda_i=0} ,\label{eq:dotwiseries}
\end{align}
with the first term in Eq.~\eqref{eq:dotqiseries} representing the heat current that is transferred from a thermal bath $i$ when $\lambda_i=0$. We notice that in the case of the coherent power the first (zero-order) term in Eq.~\eqref{eq:dotwiseries} is exactly zero, $ \Dot{W}_i|_{\lambda_i=0} = 0$, highlighting the fact that such power sources are driven by the presence of environmental coherence and vanish otherwise.

\section{Regimes of operation}\label{sec:resultsregime}

The comparison of extra terms in the energetics of the machine due to the presence of coherence in the reservoirs implies that it can function in different regimes of operation that can significantly differ from the thermal case. 

In particular, the changes in coherence of the reservoir's auxiliary units may lead to extra input (or output) work, enlarging the accessible tasks that the machine can perform. We denote the total input work as $\Dot{W}=\sum_{i=1}^3 \Dot{W}_i$, keeping the sign convention that associates the consumption of work from the reservoirs to the system with a positive sign. Analogously heat currents $\dot{Q}_i$ from the reservoir $i$ are positive when flowing from the reservoir into the system.

The possible regimes allowed in the setup are summarised in Table~\ref{table:functionings} and are explained as follows:
\begin{itemize}
\item \textbf{Quantum absorption refrigerator I}: A heat current from the hot bath ($\Dot{Q}_3>0$) to the intermediate temperature one ($\Dot Q_2 <0$), allows the removal of energy from the cold bath ($\Dot{Q}_1>0$) 
leading to refrigeration without the need of an external driving or power ($\Dot{W}=0$).
\item \textbf{Heat pump II}: inverting regime I, the machine is able to pump heat into the hot reservoir ($\Dot Q_3 <0$) while absorbing the heat current from the intermediate temperature bath ($\Dot Q_2 >0$) and dissipating a remnant amount of heat in the cold one ($\Dot Q_1 < 0$). 
\item \textbf{Power and heat driven refrigerator III}: the inclusion of coherent work allows cooling of the cold bath ($\Dot{Q}_1>0$) by means of both input work ($\Dot W>0$) and the heat flow from the hot bath ($\Dot{Q}_3>0$) to the intermediate temperature one ($\Dot Q_2 < 0$).
\item \textbf{Power and heat driven pump IV}: input work ($\Dot{W}>0$) can also assist heating of the hot reservoir ($\Dot Q_3 <0$) in combination with the heat flow from the intermediate bath ($\Dot Q_2 >0$) to the cold one ($\Dot Q_1 < 0$). 
\item \textbf{Hybrid power driven refrigerator and heat pump V}: two useful thermodynamic tasks are simultaneously performed using a single input: by consuming input work ($\Dot{W} > 0$) the machine is able to both extract heat from the cold reservoir ($\Dot{Q}_1>0$) transferring it to the intermediate reservoir (power driven refrigerator) ($\Dot{Q}_2<0$) while pumping heat to the hot reservoir (heat pump) ($\Dot{Q}_3<0$).
 \item \textbf{Hybrid heat engine and heat pump VI}: two useful thermodynamic tasks being carried out simultaneously, the heating of the hot reservoir (heat pump) ($\Dot{Q}_3<0$) and the production of work (heat engine) ($\Dot{W} <0$) powered by a heat current from the intermediate $(\Dot Q_2 > 0)$ to the cold reservoirs ($\Dot Q_1 < 0$).
\item \textbf{Dual sink accelerator VII}: the machine acts by accelerating the heat currents from the hot source ($\Dot Q_3 >0$) to the other terminals ($\Dot Q_2 <0$ and $\Dot Q_1<0$), by using  input work ($\Dot W >0$).
 \item \textbf{Triple power driven pump VIII}: the machine consumes input work ($\Dot{W}>0$) to pump heat into all three terminals ($\Dot Q_i < 0$ for all $i$).

\end{itemize}


\begin{figure*}[t!]

\subfigure{\includegraphics[height=4.8cm]{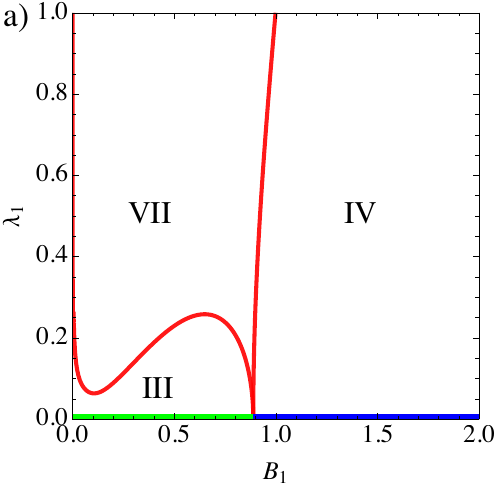}}
\subfigure{\includegraphics[height=4.8cm]{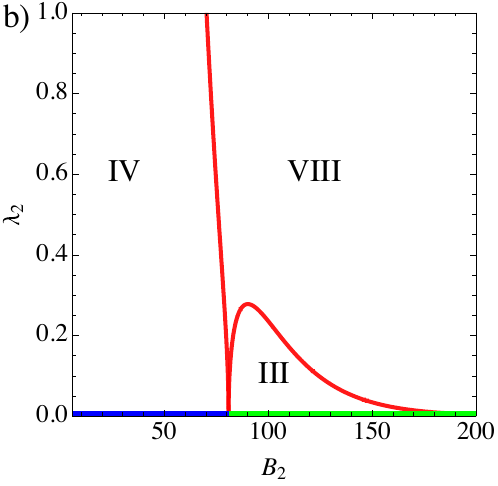}}
\subfigure{\includegraphics[height=4.8cm]{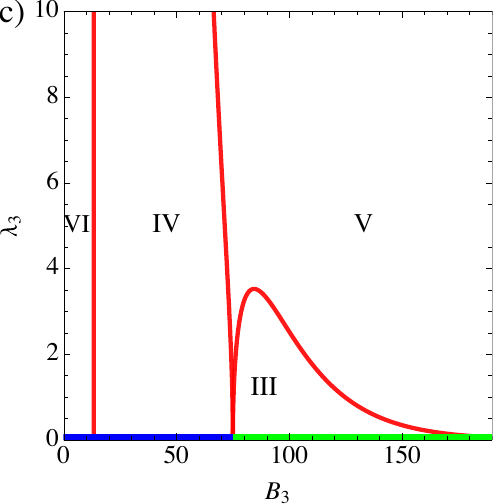}}
\hfill
\caption{Operational diagram for the functioning of the machine in the resonance case for: a) coherence in the cold bath, with $B_2 = 12.$; b) coherence in the intermediate bath with $B_1 = 6.$; and c) coherence in the hot bath with $B_1 = 6$. The green and blue horizontal lines, for $\lambda_i=0$, correspond to the regimes I and II, respectively. The red lines serve as boundaries that mark the shifts between the regimes, which are made possible by the input coherence in the environmental units. The rest of the parameters are $T_1 = 1.$, $T_2 = 6.$, $T_3 = 10.$, $\gamma_1= 8.7 \times 10^{-3}$, $\gamma_2 = 5.7 \times 10^{-3}$, and $\gamma_3 = 7.5 \times 10^{-3}$.}
\label{fig:cohbaths}
\end{figure*}

The resulting regimes of operation can be divided into the following main categories. \emph{Single task} regimes, which are characterized by the performance of only one useful thermodynamic operation achieved by input energy from a single source, as in many prototypical heat engines and refrigerators. Single task regimes include regimes I, II, and VIII. However, due to the non-equilibrium nature of the reservoirs, we obtain additional regimes where a single task is performed using multiple input resources. We refer to these as \textit{combined multisource regimes}, which include regimes III and IV. Moreover we also obtain \textit{hybrid multitask regimes} which are the ones that perform more than one useful thermodynamic task simultaneously, as in the case of regimes V and VI. Finally, we refer to regime VII to as a \emph{leaky} regime where no particularly useful task is achieved. For the above characterization we referred to useful tasks to either cooling the cold bath ($\dot{Q}_1 \geq 0$), heating the hot bath ($\dot{Q}_3 < 0$) or extracting work ($\dot{W}_i < 0$).

\subsection{Thermal collision units}\label{sec:regnocoh}

After defining the possible operational regimes of our coherent thermal machine, we briefly summarize the behavior of the machine for the thermal case. In this situation, no coherence in the reservoirs is considered, $\lambda_1=\lambda_2=\lambda_3=0$, thus the state of each bath is described by a thermal Gibbs state $\rho_R^{(i)}=e^{-\beta_i H_R^{(i)}}/Z_i$.
Due to the equilibrium nature of the reservoirs (and the absence of extra conserved quantities or external sources), there is no work contribution in this case, $\Dot{W}=0$. Therefore, we recover the standard expression for the heat flow, $\Dot{Q}_i=-\Dot{E}_R^{(i)}$, as follows from Eq.~\eqref{eq:heat}.

In the long time limit, the machine reaches a nonequilibrium steady state verifying $\Dot{\rho}_S=0$ (see Appendix ~\ref{steadystate}), which is characterised by the heat currents:
\begin{align}
\dot Q_1 &= -B_1 V_{ss}
\label{eq:Q11}
\\
\dot Q_2 &= B_2 V_{ss}
\label{eq:Q21}
\\
\dot Q_3 &= -B_3 V_{ss}
\label{eq:Q3}
\end{align}
all of them proportional to the common factor:
\begin{widetext}
    \begin{align}
    V_{ss}= \frac{4 (-\bar{n}_1 \bar{n}_3 +\bar{n}_2(1+\bar{n}_1+\bar{n}_3))\gamma_1 \gamma_2 \gamma_3}{(\bar{n}_2+\bar{n}_3+3 \bar{n}_2 \bar{n}_3)\gamma_2\gamma_3+ \gamma_1((1+2 \bar{n}_1 +2\bar{n}_2+3 \bar{n}_1 \bar{n}_2)\gamma_2 +(1+2 \bar{n}_1+\bar{n}_3+3 \bar{n}_1 \bar{n}_3)\gamma_3)}.
\end{align}
\end{widetext}
The equilibrium condition for which all heat fluxes become zero and no entropy is produced, is hence verified for $V_{ss}=0$, leading to: 
\begin{equation}\label{eq:btransition}
\frac{B_1}{T_1} - \frac{B_2}{T_2}+ \frac{B_3}{T_3}=0,
\end{equation}
where we recall that $B_3 = B_2 - B_1$. The above relation provides the transition point between the two possible regimes of operation: the machine operates as a quantum absorption refrigerator I when $B_1(\beta_1 + \beta_3) \leq B_2(\beta_3 + \beta_2)$ and as a heat pump II for $B_1(\beta_1 + \beta_3) \geq B_2(\beta_3 + \beta_2)$. These features are actually generic under endoreversible conditions stemming from the fact that each different transition in the system is connected to only a single reservoir, see e.g.~\cite{brunner:2012,Correa14,correa2014quantum,Hewgill_PRE2020}. While zero power is obtained when considering thermal reservoirs units colliding with the system, the situation will radically change when the auxiliary units of the baths are endowed with some amount of coherence.

\subsection{Coherent collision units}\label{sec:regcoh}

We now inject some small amount of coherence in one of the baths while the other two are kept thermal. We consider three separate cases: when coherence is either in the cold bath ($\lambda_1 \neq 0$ and $\lambda_2=\lambda_3=0$), in the intermediate temperature bath ($\lambda_2 \neq 0$ and $\lambda_1=\lambda_3=0$) or in the hot bath ($\lambda_3 \neq 0$ and $\lambda_2=\lambda_1=0$), respectively.

To evaluate the regimes, the work coming from the environmental auxiliary units of the reservoir that contains coherence needs to be taken into account; this work term $\dot W_{i}$ depends on $\lambda_i$, as it follows from Eq.~\eqref{eq:coherentwork}. Due to the changes in the coherence of the auxiliary units during the collisions, new additional regimes of operation, besides the ones found for the thermal case, appear. The coherent work and the heat current stemming from the coherent reservoir auxiliary units follow from Eqs.~\eqref{eq:coherentwork} and \eqref{eq:heat} and are explicitly calculated in Appendix \ref{ap:power}, whilst heat coming from the other thermal baths is simply given by $\Dot{Q}_j=-\Dot{E}_j$, as it corresponds to the thermal case.

\subsubsection{Coherence in the cold reservoir} \label{sec:cohcold}
Assuming some initial coherence in the cold environmental units, we find that our setup is able to operate in regimes III, IV and VII [see Table~\ref{table:functionings}] by modulating the parameters and the amount of injected coherence in the reservoir.

In Fig.~\ref{fig:cohbaths}a) we show a diagram for the functioning of the machine in terms of the units energy spacing $B_1$ and the coherence strength parameter $\lambda_1$. When $\lambda_1=0$ (horizontal axis of the plot) and while varying $B_1$, we obtain the standard absorption refrigerator I for $\beta_1 B_1 < (\beta_2 B_2 - \beta_3 B_3)$, and the heat pump II associated with the thermal case for $\beta_1 B_1 > (\beta_2 B_2 - \beta_3 B_3)$. However as soon as $\lambda_1$ takes non-zero values, regime I is then replaced by the combined multisource regime III which, apart from the heat current, consumes  a positive amount of coherent work for chilling the cold bath. By further increasing $\lambda_1$ the regime III transforms into the dual sink accelerator VII, where spontaneous heat currents are accelerated by input power, without performing any particularly useful thermodynamic task. When $\beta_1 B_1 > (\beta_2 B_2 - \beta_3 B_3)$, the heat currents become inverted and the machine transitions from regimes III-VII to the combined multisource regime IV pumping heat into the hot reservoir. The analytical values of $\lambda_1$ at which this transition occurs are given in Appendix~\ref{App:lambda}. We find that, in all three regimes above, the coherent work is positive. Additionally, coherence not only enables new modes of operation but acts as a driving fuel for thermodynamics processes even in situations where there is no temperature difference (see Appendix ~\ref{Ap:sametemp}).

\subsubsection{Coherence in the intermediate temperature reservoir}\label{sec:cohintermediate}

When coherence is added to the intermediate reservoir, we find a similar situation, with the combined multisource regimes III and IV, but instead of the dual sink accelerator VII, the triple power driven pump VIII becomes possible, as illustrated in Fig.~\ref{fig:cohbaths}b). 

As before, in the absence of coherence (horizontal axis) we recover the standard absorption regimes for heat pumping II when $\beta_2 B_2 < (\beta_1 B_1 + \beta_3 B_3)$, and refrigeration I when $\beta_2 B_2 > (\beta_1 B_1 + \beta_3 B_3)$, which transform into the combined multisource regimes IV and III respectively as long as some initial coherence is introduced $\lambda_2 \neq 0$ acting as an extra source of power $\dot{W}_2 > 0$ assisting the corresponding tasks. We notice the mirror symmetry between Figs.~\ref{fig:cohbaths}a) and ~\ref{fig:cohbaths}b), where also regime III is lost when greater amounts of initial coherence are considered. In this case, the machine switches from either heating up only the hot reservoir, regime III, or absorbing heat from the cold one, regime IV, to a configuration pumping heat simultaneously to the three terminals, regime VIII. Notice however, that in this case the machine still performs a useful thermodynamic task, that is, it still pumps heat into the hot reservoir, which is driven by input coherent work $\dot{W} > 0$. The values of $\lambda_2$ at which the transition between regimes occur can be found in Appendix~\ref{App:lambda}.

\subsubsection{Coherence in the hot reservoir}\label{sec:cohot}

As for the last case, we investigate the functioning of the machine when coherence is injected into the hot reservoir. In this case, we report the appearance of two other regimes corresponding to hybrid multitask configurations, namely, V and VI, apart from the combined multisource regimes III and IV obtained before [see Table~\ref{table:functionings}].

The regimes are plotted in Fig.~\ref{fig:cohbaths}c) as a function of $B_3$ and $\lambda_3$. In this case, when $\beta_3 B_3 < (\beta_2 B_2 - \beta_1 B_1)$, leading to the absorption heat pump II in the absence of coherence, the heat fluxes keep the same sign, but the coherent work can become negative. Indeed, for small energy spacing between levels $\ket{1}$ and $\ket{2}$ such that $\beta_2 B_3 < -\beta_2 B_1+\ln\left[\frac{e^{\beta_1B_1}\gamma_2 (\gamma_1-\gamma_3)+\gamma_3(\gamma_1+\gamma_2)}{\gamma_1(\gamma_2+\gamma_3)}\right]$, we obtain $\Dot{W}_3 <0$ and hence the machine executes another task besides heating up the hot reservoir: it also delivers work at the same time. Thus, it is noteworthy that this device is capable of carrying out more than one beneficial tasks at the same time, highlighting the fact that incorporating coherence may enable the execution of hybrid operations. 

Moreover, for sufficiently large $B_3$ and $\lambda_3$, we find a second hybrid regime, namely regime V, acting as a power-driven refrigerator and heat pump simultaneously. The change from the multisource combined refrigeration regime III to V may be enabled for larger $B_3$ when sufficient coherence is introduced in the reservoir units, such that a certain transition point in the values of $\lambda_3$ is reached (see Appendix~\ref{App:lambda}).

Finally, we remark that by inclusion of coherence in the auxiliary units of a single thermal reservoir, all the possible regimes in Table~\ref{table:functionings} become possible for certain regions of parameters, including both hybrid multitask regimes and combined multisource ones.

\begin{figure*}[t!]
\subfigure{\includegraphics[height=3.1cm]{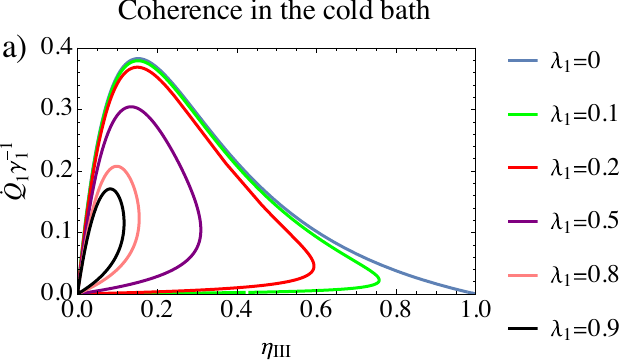}}
\subfigure{\includegraphics[height=3.1cm]{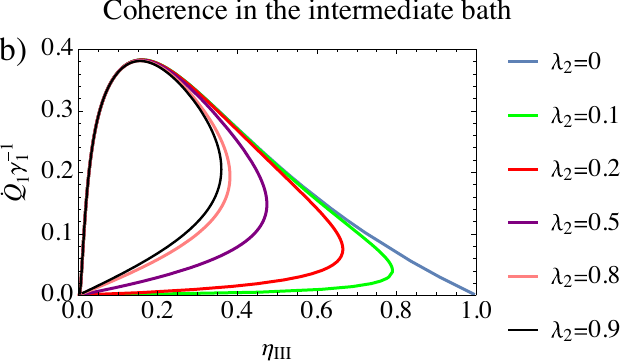}}
\subfigure{\includegraphics[height=3.1cm]{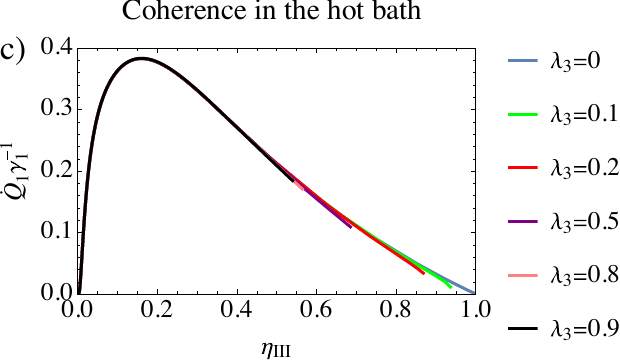}}
\hfill
\caption{Quantifying the performance of the machine in regime III via its efficiency $\eta_{III}$ in terms of the cooling power $\Dot{Q}_1$ in units of $T_1 \gamma_1$  for when coherence is in a) the cold bath, with $B_2 = 9.5$; b) the intermediate temperature bath $B_1 = 1.24$; c) the hot bath with $B_1 = 1.24$. The rest of the parameters are $T_1 = 1.$, $T_2 = 2.$, $T_3 = 60.$, $\gamma_1= 8.7 \times 10^{-3}$,$\gamma_2 = 5.7 \times 10^{-3}$, and $\gamma_3 = 7.5 \times 10^{-3}$.}

\label{fig:cohfridge}
\end{figure*}

\section{Performance of the machine}\label{sec:performance}
Generally, the functioning of a quantum thermal machine performing a single thermodynamic task is characterised by a generic thermodynamic efficiency as the ratio of the device's useful output to the energy cost or input. For a multitasking thermal device coupled to several reservoirs at different temperatures as well as for regimes where multiple inputs are allowed, it is crucial to define a meaningful efficiency in a comprehensive and standardized approach.
Such efficiency cannot treat all energy fluxes on an equal footing, but it needs to take into account the fact that, for instance, cooling a cold reservoir is more useful, thermodynamically speaking, than cooling another reservoir with a higher temperature. In Ref.~\cite{manzano2020hybrid} a universal efficiency based on nonequilibrium free energy contributions has been proposed to handle such situations:
\begin{equation}\label{eq:eta_reg}
    \eta= \frac{-\sum^{-}_{\alpha}\Dot{W}_{\alpha} + \sum^{+}_{i} \Dot{Q}_i (\frac{T_r}{T_i}-1)}{\sum^{+}_{\alpha}\Dot{W}_{\alpha} - \sum^{-}_{i} \Dot{Q}_i (\frac{T_r}{T_i}-1)},
\end{equation}
where the numerator represents the useful contributions and processes (output) and the denominator encompasses all wasteful energy contributions (input), and we have defined $\sum^{\pm}_{i} x_i=(x_i \pm |x_i|)/2$.  We remark that the above formula is slightly different than the one reported in \cite{manzano2020hybrid} due to the fact that here we used a different sign convention for work. In agreement with the second law of thermodynamics, Eq.~\eqref{eq:eta_reg} is upper bounded by unity ($0 \le \eta \le 1$) which means that the machine attains reversibility at maximum efficiency while running infinitely slow.

Importantly, Eq.~\eqref{eq:eta_reg} depends on a reference temperature $T_r$ which serves as a benchmark for evaluating the practicality of a certain thermodynamic task. Its choice is arbitrary and highlights an intrinsic freedom in characterizing the available resources. Different choices of $T_r$ may lead to recover different known expressions for the efficiency in standard devices. For instance, when the machine operates as a heat engine ($\dot W<0$) coupled to two reservoirs with $T_1<T_2$, Eq.~\eqref{eq:eta_reg} becomes, for $T_r=T_1$,
\begin{equation}\label{eq:engine}
    \eta_E=\frac{1}{\eta_C} \frac{-\Dot{W}}{\Dot{Q}_2} \leq 1,
\end{equation}
with $\eta_C=1-\frac{T_1}{T_2}$ denoting the Carnot efficiency. In this case, the choice of $T_r$ is attributed to the typical situation in traditional steam engines, where $T_1$ is the ambient temperature and hence taken for granted, while a heat flux from a hotter source  $T_2$ is a useful resource that requires burning some fuel. However, taking $T_r = T_2$ would lead to an equally meaningful alternative definition for the efficiency of heat engines, $\eta_E^\prime = \dot{W}/(\frac{T_2- T_1}{T_1} \dot{Q}_1)$ remarking the fact that dissipating heat into a low temperature sink is an equally useful resource~\cite{manzano2020hybrid}.

In our scenario, we are interested in assessing the performance of the three-level machine in contact with one coherent reservoir and two thermal ones. This configuration can be seen as a three-terminal device whose modes of operation have an associated efficiency that can be evaluated by using Eq.~\eqref{eq:eta_reg}. In the following, we will assign the reference temperature to either the intermediate reservoir $T_r=T_2$ or the cold one $T_r = T_1$ depending on the situation. By making this particular choice, we are able to identify useful processes such as cooling the cold bath and heating the hot one by using different available resources, and recover all the standard definitions of efficiency in the thermal case~\cite{KosloffARPC2014,brunner:2012,Levy2012}.

In the following, we will define the efficiency of the regimes using combined resources to perform a single task, i.e. regimes III and IV, in the first place, and then proceed to discuss the hybrid multitask regimes V and VI. We remark that while in Fig.~\ref{fig:cohbaths} we used specific values of parameters that help illustrate the regimes of operation of the machine, here in order to better assess our device's functionality, we explored a broad range of parameters for the temperatures of the reservoirs and energy levels of the system. This allowed us to identify the specific parameters that enable the best description of the regimes' optimal performance.

\subsection{Performance of multisource combined regimes}\label{sec:HP}

\subsubsection{Power and heat driven refrigerator III}
As commented above, in this regime cooling of the cold reservoir ($\Dot{Q}_1>0$) is powered by both input coherent work ($\Dot{W}_i>0$) and the natural heat flow from the hot reservoir ($\Dot{Q}_3>0$) to the intermediate one ($\Dot{Q}_2 < 0$). This means that we have two combined resources that drive the refrigerator to function. The corresponding efficiency from Eq.~\eqref{eq:eta_reg} with $T_r = T_2$ hence reads
\begin{equation}
    \eta_{III}=\frac{\Dot{Q}_1}{\epsilon_{AR}^{\max}\Dot{Q}_3 + \epsilon_{R}^{\max} \Dot{W}},
\end{equation}

with $\epsilon_{AR}^{\max} :=\frac{T_1(T_3-T_2)}{T_3(T_2-T_1)}$ the Carnot coefficient of performance (COP) of the quantum absorption refrigerator~\cite{Levy2012,brunner:2012,Correa14} and $\epsilon_{R}^{\max} := \frac{T_1}{T_2-T_1}$ the Carnot COP for a standard power-driven fridge in a two-terminal configuration~\cite{manzano2020hybrid}. The above expression highlights the different accounting of the two input energy sources (heat and work), leading to different maximum COP ($\epsilon_{AR}^{\max}$ and $\epsilon_{R}^{\max}$), when the other source is absent. 

Fig.~\ref{fig:cohfridge} shows the cooling power $\Dot{Q}_1$ versus the efficiency $\eta_{III}$ for different values of $\lambda_i$, when coherence is present in the cold, intermediate and hot reservoir auxiliary units, respectively.  Notably, the role of coherence in the performance of the three-terminal machine can be observed from the different curves associated to different values of $\lambda_i$.

For the cold reservoir, Fig.~\ref{fig:cohfridge}a), we observe that both the cooling power and the efficiency decrease rapidly when increasing $\lambda_1 $ in comparison with the thermal case (blue line). In this case, the maximum cooling power $\Dot{Q}_{1}^{\rm max}$ is highly reduced together with its efficiency. It is also worth noticing that the maximum efficiency cannot achieve the reversible point in the presence of coherence anymore, $\eta_{III}^\mathrm{max} < 1$, and hence it no longer leads to zero currents; in other words, endoreversibility is lost.

In Fig.~\ref{fig:cohfridge}b), where coherence is added to the intermediate temperature reservoir, a slightly distinct pattern is observed. There for low efficiencies ($B_3 \gtrsim 10.$), the values of the cooling power remain closely similar to their thermal counterparts when varying $\lambda_2$, including the values of the maximum cooling power $\dot{Q}_1^{\max}$. However, when going to higher efficiencies ($B_1 <B_2 \lesssim 10.$), we observe again a reduction in power and efficiency with respect to the thermal case as $\lambda_2$ increases. The maximum efficiency points for non-zero values of coherence are clearly reduced where $\eta_{III}^{\rm max}$ falls below $0.4$ for $\lambda_2 > 0.5$.

In Fig.~\ref{fig:cohfridge}c), corresponding to coherence added into the hot reservoir auxiliary units, the machine power and efficiency does not significantly change with respect to the baseline curve associated to the thermal case. However, we see that the curves are cut for large values of the efficiency, since that values would be forbidden within regime III. 

The above results show that, even if coherence may act as an extra resource for refrigeration in this regime, in practice it is unable to improve the performance of the machine in any case. That is, it does not contribute to achieve either larger cooling powers or efficiencies as compared to the thermal case, but just increases the wasteful dissipation of heat into the intermediate temperature reservoir $\dot{Q}_2$. Moreover, when coherence is injected in the cold reservoir it becomes especially detrimental to the function of the machine producing a reduction of both power and efficiency.

\begin{figure*}[bht]

\subfigure{\includegraphics[height=3.2cm]{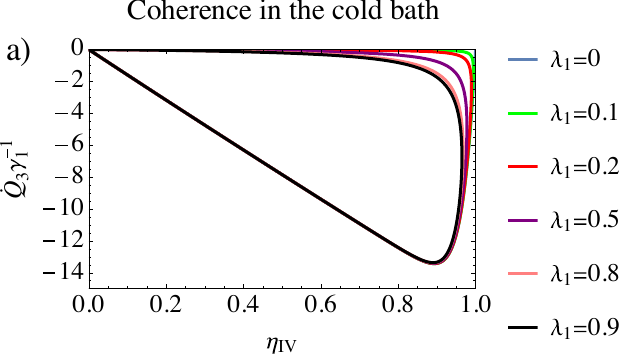}}
\subfigure{\includegraphics[height=3.2cm]{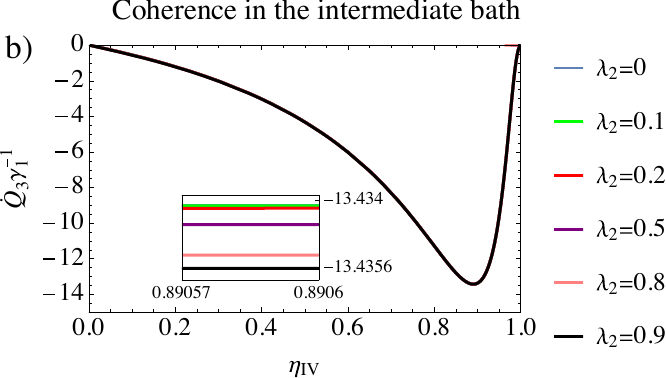}}
\subfigure{\includegraphics[height=3.2cm]{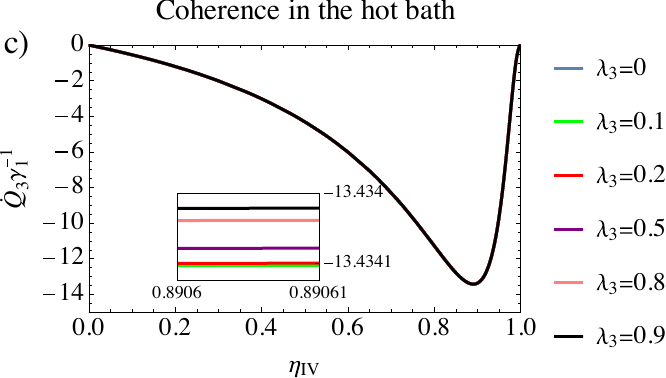}}
\hfill
\caption{Quantifying the performance of the machine in regime IV via its efficiency $\eta_{IV}$ in terms of the heat flow from the work bath $\Dot{Q}_3$ in units of $T_1 \gamma_1$ for when coherence is in a) the cold bath, with, $B_2 = 35.31$; b) the intermediate temperature bath, with $B_1 = 4.34$; and c) the hot bath, with $B_1 = 4.34$. The rest of the parameters are  $T_1 = 1.$, $T_2 = 30.$, $T_3 = 60.$, $\gamma_1= 8.7 \times 10^{-3}$,$\gamma_2 = 5.7 \times 10^{-3}$, and $\gamma_3 = 7.5 \times 10^{-3}$. }
\label{fig:cohperformance}
\end{figure*}

\subsubsection{Power and heat driven pump IV}
In this regime, the heating of the hot bath ($\Dot{Q}_3<0$) is achieved by combining two input sources: the input coherent work ($\Dot{W}_i>0$) and input heat from the intermediate reservoir ($\dot{Q}_2 > 0$) which is dissipated into the cold bath ($\Dot{Q}_1<0$). The efficiency that illustrates this concurrent operation using $T_r = T_1$ is provided by
\begin{equation}\label{eq:etahp}
    \eta_{IV}= \frac{- \Dot{Q}_3}{\eta_{AP}^{\max} \Dot{Q}_2+ \eta_P^{\max} \Dot{W}},
\end{equation}
where $\eta_{AP}^{\max} = \frac{T_3}{T_2}\frac{(T_2 - T_1)}{(T_3 - T_1)}$ is the Carnot efficiency for the standard absorption heat pump~\cite{brunner:2012} and $\eta_P^{\max} = T_3/(T_3 - T_1)$ corresponds to the Carnot efficiency for power-driven heat pumping in a two-terminal configuration. In analogy to the previous multisource regime III, the analysis of the efficiency $\eta_{IV}$ reveals the combination of standard resources for heat pumping, mixed with the corresponding weights.

In Fig.~\ref{fig:cohperformance}, we plot the heating power $\Dot{Q}_3$ against the efficiency $\eta_{IV}$ for different values of $\lambda_i$, when coherence is in the cold, intermediate and hot reservoir units, respectively.

For the case of the cold reservoir, Fig.~\ref{fig:cohperformance}a), we see that when $\eta_{IV}\lesssim 0.85$, incorporating coherence into the cold bath does not affect significantly the heating power, which remarkably achieves its maximum close to the maximum efficiency limit. However, as the value of $\lambda_1$ increases, there is a decrease in 
the maximum achievable efficiency, which reduces to $0.9$ for $\lambda_1 = 0.9$. Moreover there is a small decrease in the maximum power which reduces from $|\Dot{Q}_{3}^{\rm max}|=13.4329$ for $\lambda_1=0$, to $13.3713$  for $\lambda_1=0.9$.

On the other hand, when coherence is added to either the intermediate temperature reservoir, see Fig.~\ref{fig:cohperformance}b), or hot reservoir, see Fig.~\ref{fig:cohperformance}c), the curves remain practically unaltered with respect to the thermal case. The values of the maximum heating power can be however slightly increased for coherence in the intermediate reservoir as highlighted in the inset graph of Fig.~\ref{fig:cohperformance}b). On the contrary, in the case of coherence in the hot reservoir the maximum  heating power $|\Dot{Q}_{3,\rm max}|$ slightly decreases for increasing values of $\lambda_3$ as shown in the inset of Fig.~\ref{fig:cohperformance}c).


\subsection{Performance of the hybrid multitask regimes}\label{sec:hybrid}

\begin{figure*}[bht!]

\subfigure{\includegraphics[height=3.1cm]{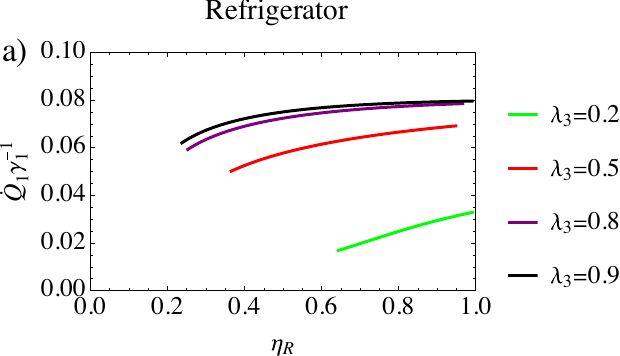}}
\subfigure{\includegraphics[height=3.1cm]{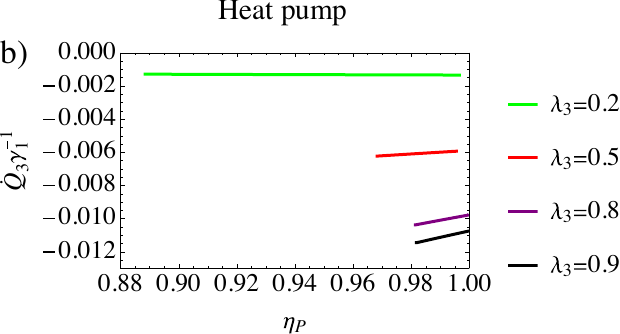}}
\subfigure{\includegraphics[height=3.1cm]{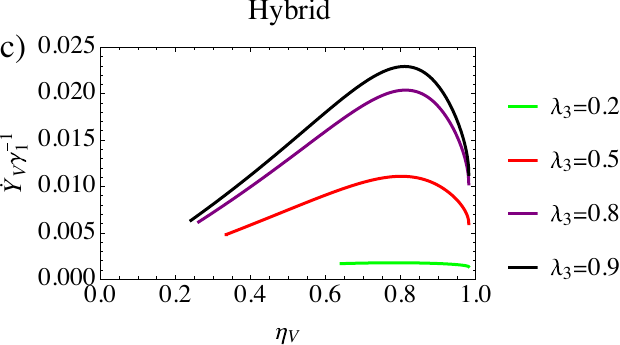}}
\hfill
\caption{Quantifying the performance of the machine in regime V for different values of $\lambda_3$ in the case of coherence in the hot bath for the separate output operation of both thermodynamic tasks: the power-driven refrigerator a) in terms of the cooling power $\Dot{Q}_1$ vs its efficiency $\eta_R$; the power-driven pump b) in terms of the heating power $\Dot{Q}_3$ vs its efficiency $\eta_P$; the whole hybrid process c) in terms of the combined output tasks $\Dot{Y}_{V}$ which is the numerator of Eq.~\eqref{eq:hybrid5} vs the efficiency of the hybrid machine $\eta_{V}$, all heat fluxes and total output power are in units of $T_1 \gamma_1$. The parameters are $T_1=1.$, $T_2 = 1.1$, $T_3 = 60.$, $B_1=4.34$, $\gamma_1= 8.7 \times 10^{-3}$,$\gamma_2 = 5.7 \times 10^{-3}$, and $\gamma_3 = 7.5 \times 10^{-3}$.}
\label{fig:cohybrid}
\end{figure*}

\subsubsection{Hybrid power driven refrigerator and heat pump V}
We now analyze the performance of the hybrid regime performing refrigeration of the cold reservoir ($\dot{Q}_1 > 0$) and heating of the hot one ($\dot{Q}_3 < 0$) simultaneously, while driven by input coherent work ($\dot{W} > 0$). As shown in Fig.~\ref{fig:cohbaths}c, this regime V can only occur when coherence is introduced to the units of the hot reservoir. The efficiency of this hybrid process for $T_r = T_2$ is given by:
\begin{equation}\label{eq:hybrid5}
\eta_{V} = \frac{\dot{Q}_1(\frac{T_2}{T_1}-1) + \Dot{Q}_3(\frac{T_2}{T_3}-1)}{\Dot{W}_3}, 
\end{equation}
which can be split into two contributions $\eta_{V}= \eta_R+\eta_{P}$ corresponding to the separate operation of both tasks:
\begin{align}
    \quad \eta_R= \frac{\Dot{Q}_1}{\epsilon_R^{\max} \Dot{W}_3} \quad \text{,} \quad \eta_{P}=\frac{- \Dot{Q}_3}{\eta_P^{\max} \Dot{W}_3},
\end{align}
where the subscript $R$ stands for refrigeration and $P$ for pump. Here $\eta_R$ is the normalized efficiency of a power-driven refrigerator, whose maximum value corresponds to reaching the Carnot COP $\epsilon_R^{\max}=T_1/(T_2 - T_1)$. On the other side, $\eta_P$ is the normalized efficiency for a power-driven heat pump, maximized for the corresponding Carnot efficiency $\eta_P^{\max} = T_3/(T_3 - T_2)$.

 The three plots in Fig.~\ref{fig:cohybrid} show power-efficiency diagrams for the operating regime V when coherence is present in the hot reservoir, for different values of $\lambda_3$. We plot both the diagrams for the two individual tasks being performed, namely, power-driven refrigeration, Fig.~\ref{fig:cohybrid}a), and power-driven pumping of heat, Fig.~\ref{fig:cohybrid}b), as well as a diagram for the overall hybrid operation, Fig.~\ref{fig:cohybrid}c), where the total output power $\Dot{Y}_V := \dot{Q}_1(\frac{T_2}{T_1}-1) + \Dot{Q}_3(\frac{T_2}{T_3}-1)$ is plotted as a function of hybrid efficiency $\eta_V$.

In Fig.~\ref{fig:cohybrid}a), the cooling power $\Dot{Q}_1$ is plotted against the refrigeration COP $\eta_{R}$ within the range of parameters leading to regime V. In contrast to the previous combined regimes, here we see a noticeable enhancement in both the cooling power as the value of $\lambda_3$ increases for similar efficiency values, such that its maximum value remains close to unity. A similar pattern is also observed in Fig.~\ref{fig:cohybrid}b) for the heating power as a function of the corresponding efficiency $\eta_P$: adding coherence improves greatly the heating of the hot reservoir while keeping high efficiencies. As a result, when combining the performance of the two tasks Fig.~\ref{fig:cohybrid}c), we see that the hybrid machine shows similar enhancements in power in regions with high efficiencies. In that case, when increasing coherence we can clearly identify a peak in the hybrid power $Y_V$ corresponding to high efficiencies at maximum power around $0.8$.


\subsubsection{Hybrid heat engine and pump VI}
Finally, we determine the functionality of the hybrid heat engine and pump VI. This regime is only available when coherence is injected in the hot reservoir and comprises the performance of two useful thermodynamics tasks simultaneously, namely, heating of the hot reservoir ($\Dot{Q}_3<0$) and generating work ($\Dot{W}_3 < 0$) using a heat current from the intermediate ($\dot{Q}_2 > 0$) to the cold reservoirs ($\dot{Q}_1 < 0$). The corresponding efficiency for $T_r = T_1$ is given by
\begin{equation}\label{eq:hybrid6}
    \eta_{VI}= \frac{-\Dot{W}_3 + \Dot{Q}_3(\frac{T_1}{T_3}-1)}{\Dot{Q}_2(1-\frac{T_1}{T_2})},
\end{equation}
which again can be split into two contributions stemming from the two tasks being simultaneously performed, $\eta_{VI}= \eta_E + \eta_{AP}$ with:
\begin{align}
     \eta_E=\frac{- \Dot{W}_3}{\eta_C \Dot{Q}_2} \quad \text{,} \quad \eta_{AP}=\frac{- \Dot{Q}_3}{\eta_{AP}^{\max}  \Dot{Q}_2},
\end{align}
in term of standard Carnot efficiencies for a heat engine $\eta_C= 1-T_1/T_2$ and absorption pump $\eta_{AP}^{\max}$, respectively, as given above.

\begin{figure*}[!bht]
\subfigure{\includegraphics[height=3.1cm]{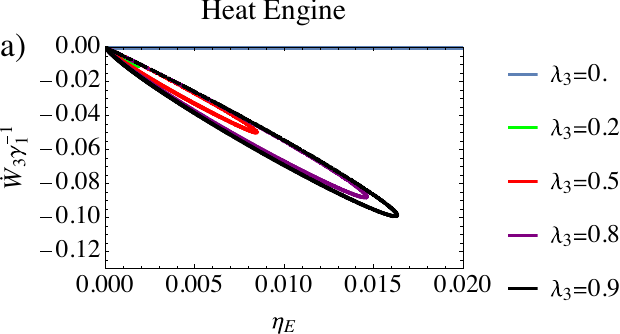}}
\subfigure{\includegraphics[height=3.2cm]{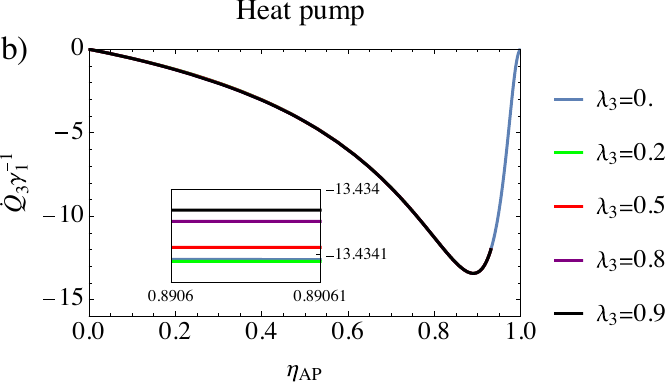}}
\subfigure{\includegraphics[height=3.2cm]{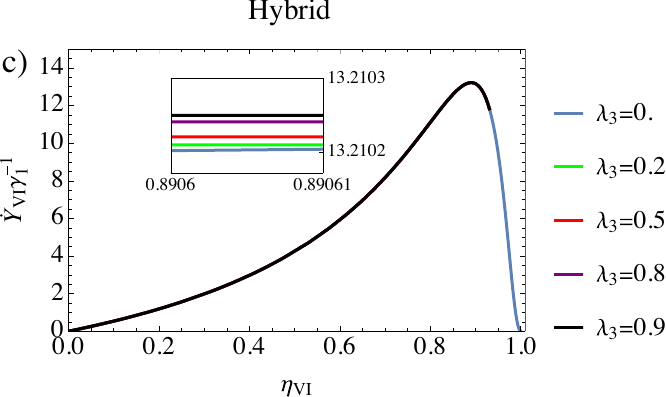}}
\hfill
\caption{Quantifying the performance of the machine in regime VI for different values of $\lambda_3$ in the case of coherence in the hot bath for the separate output operation of both thermodynamic tasks: the heat engine a) in terms of the coherent work $\Dot{W}_3$ vs its efficiency $\eta_E$;  the absorption pump b) in terms of the heating power $\Dot{Q}_3$ vs its efficiency $\eta_{AP}$; the whole hybrid process c) in terms of the combined output tasks $\Dot{Y}_{VI}$ which is the numerator of Eq.~\eqref{eq:hybrid6} vs the efficiency of the hybrid machine $\eta_{VI}$. The coherent work, the heating power and the total output power are all in units of $T_1 \gamma_1$. The parameters are $T_1=1.$ $T_2 = 30.$, $T_3 = 60.$, $B_1=4.34$, $\gamma_1= 8.7 \times 10^{-3}$,$\gamma_2 = 5.7 \times 10^{-3}$, and $\gamma_3 = 7.5 \times 10^{-3}$.}
\label{fig:cohybrid1}
\end{figure*}

Fig.~\ref{fig:cohybrid1} shows three power-efficiency diagrams for operating regime VI when coherence is present in the hot reservoir, for both individual and collective performance evaluation. The heat engine, absorption heat pump and the overall hybrid machine VI performance diagrams are plotted in Figs.~\ref{fig:cohybrid1}a), ~\ref{fig:cohybrid1}b), and ~\ref{fig:cohybrid1}c), respectively.

The heat engine performance is shown in Fig.~\ref{fig:cohybrid1}a), where
the coherent work $\Dot{W}_3$ is plotted against the engine's standard efficiency $\eta_{E}$ for different values of $\lambda_3$. The optimal configuration of the engine is reached for $|\Dot{W}^{\rm max}_3|\approx 0.1$ with its corresponding efficiency $\eta_{E}(|\Dot{W}^{\rm max}_3|)\approx 0.0162$, which actually coincides with the maximum achievable efficiency $\eta_E^{\rm max}$.

Despite the fact that coherence enables work extraction, this regime occurs at quite low efficiencies. Moreover, the effect of the coherence on the functioning of the absorption pump is not promising. By plotting $\Dot{Q}_3$ as a function of $\eta_{AP}$ in Fig.~\ref{fig:cohybrid1}b), we show that coherence does not improve power or efficiency for the absorption pump with respect to the thermal case. The heating power is slightly reduced, as emphasised in the inset. Moreover reaching efficiencies over $0.9$ is no longer possible when using coherence. 

Finally in Fig.~\ref{fig:cohybrid1}c), the total output power $\Dot{Y}_{VI}:=-\Dot{W}_3 + \Dot{Q}_3(\frac{T_1}{T_3}-1)$ of the hybrid machine is plotted with respect to its hybrid efficiency $\eta_{VI}$. This reveals that the power in this regime is slightly enhanced by the input environmental coherence, as can be better appreciated in the inset of Fig.~\ref{fig:cohybrid1}c). There, the efficiency at maximum power $\eta_{VI}(|\Dot{Q}_3^{\rm max}|) \approx 0.8906$ remains approximately constant for all values of $\lambda_3$ and efficiencies over $0.9$ cannot be reached as before.

\raggedbottom
\section{Summary and Conclusions}
\label{sec:conclusion}
In this work, our central objective has been to explore the impact of introducing an infinitesimal amount of coherence into collisional reservoirs on the performance of a three-terminal autonomous thermal machine.
Our analysis revealed the presence of new operational modes enabled by the presence of coherence, and we evaluated the performance of all possible regimes at steady-state conditions. 
The amount of coherence, the energy spacing in the different transitions of the machine and the temperatures of the reservoirs play a crucial role in determining the functionality of our device. In this context, we used a broad range of parameters to highlight the versatility of our machine. 

Our results showed that the presence of quantum coherence can lead to an extra source of work that may be combined with heat currents from different terminals to operate both multisource combined regimes performing a single thermodynamic task, and hybrid multitask regimes, for which two tasks are operated in parallel. The multisource combined regimes involve two resources of energy, namely, heat and work to perform a single task such as cooling (regime III) or heat pumping (regime IV). On the other hand, the hybrid regime V performs both refrigeration of a cold reservoir and heat pumping into a hot one from input coherent work, and regime VI extracts work and pumps heat into a hot reservoir simultaneously by using only a heat current. 

Based on our findings, it becomes evident that, for the combined multitask regime III, coherence does not provide any significant enhancement to the overall operation. In other words, achieving a greater cooling power and/or efficiency is not possible in any scenario in the presence of coherence when compared to the standard thermal case. In the case of the multisource pump IV, the efficiency at maximum power $\eta_{IV}(|\Dot{Q}_3^{\rm max}|)$ is close to unity for any situation of coherence in the reservoirs unlike in regime III. Although there is no noticeable improvement in the heating power due to coherence in hot reservoirs, there is a slight increase in $|\Dot{Q}_3^{\rm max}|$ as the value of $\lambda_2$ becomes larger when coherence is in the intermediate temperature reservoir.

As far as the hybrid regimes are concerned, we find that the global hybrid efficiencies for regimes V and VI can be split into separate efficiencies that describe each specific output thermodynamic task. Increasing the amount of coherence in the hot bath leads to the enhancement of the performance of both regimes V and VI. For instance, the power-driven refrigerator and power-driven pump in regime V show improvement in terms of the cooling and heating power, for fixed values of their individual efficiencies and its global hybrid one. We have also shown that the operation of regime VI is slightly enhanced due to coherence, while its impact on the individual thermodynamic operations yielded ambivalent  results. The engine regime manifests enhancements of both maximum coherent work and its efficiency at maximum power, whereas in the absorption pump regime, there is a slight decrease of the maximum heating power.

In summary, when we compare between both types of regimes, it becomes apparent that coherence has a more advantageous impact on the performance of the hybrid operations of the machine in terms of power for fixed efficiency, as opposed to the multisource combined machines.
The results reported here might be compared with previous results that did not take into account the split of the energy exchanged with the nonequilibrium reservoir in work and heat contributions. 

Possible follow up work may also consider the combination of different coherence sources in the same setup, or the performance of coherent reservoirs in energy-harvesting setups with more conserved quantities such as thermoelectric devices~\cite{Sothmann2015}.
It would also be interesting to analyze the fluctuations of heat and work currents in this setup model in relation to the thermodynamic uncertainty relation (TUR)~\cite{UdoPRL2015,Gingrich16}. We expect that reduced fluctuations on the coherent power may lead to TUR violations witnessing quantum-mechanical enhancements in the performance of the machine, similarly to what has been found in some thermoelectric devices~\cite{Ptaszy2018,Liu19,Lopez2023,Manzano23} and other thermal machine models~\cite{Kalaee21,Mitchison21,daSilva22} using coherent driving sources.

The supporting data for this article are openly available
from the \cite{kenza6_2024}.

\hfill

\acknowledgements
We thank Luis A. Correa and Philipp Strasberg for interesting discussions. KH and GDC acknowledge the support by the UK EPSRC EP/S02994X/1 and the Royal Society IEC\textbackslash R2\textbackslash 222003. G.M. acknowledges funding from the ’Ram\'on y Cajal’ program (RYC2021-031121-I) the CoQuSy project (PID2022-140506NB-C21) and support from the ’María de Maeztu’ project CEX2021-001164-M, funded by MICIU/AEI/10.13039/501100011033 and European Union NextGenerationEU/PRTR.

\appendix

\section{Steady state of the thermal units of the baths}\label{steadystate}
After an infinite number of collisions, the system reaches a non-equilibrium steady state (NESS) which satisfies $\dot\rho_S=0$. It can be expressed in the following form
\begin{align}\label{ness}
\rho_S^{NESS}=\begin{pmatrix}
\Tilde{\rho}_{1,1} & \Tilde{\rho}_{1,2} & \Tilde{\rho}_{1,3}\\
\Tilde{\rho}_{2,1} & \Tilde{\rho}_{2,2} & \Tilde{\rho}_{2,3}\\
\Tilde{\rho}_{3,1} & \Tilde{\rho}_{3,2} & \Tilde{\rho}_{3,3}
\end{pmatrix}	
\end{align}
The expressions corresponding to the entries of $\rho_S^{NESS}$ depend on $\lambda_i$ and can be found analytically for each particular case of coherence injected in one of the baths. 
When $\lambda_{i}=0$ with $\{i=1,2,3\}$, we find the analytical expression of the entries of the steady state density matrix~\eqref{ness} for the thermal case: 
\begin{align}
     ~~\Tilde{\rho}_{1,1} &= \frac{1}{N}(\bar{n}_3\gamma_2 \gamma_3(1+\bar{n}_2)
     \\ \nonumber
     &~~~~+ \gamma_1(1+\bar{n}_1)[(1+\bar{n}_2)\gamma_2+(1+\bar{n}_3)\gamma_3])\\ \nonumber
    \Tilde{\rho}_{2,2}&=\frac{1}{N}(\bar{n}_1\gamma_1 \gamma_2(1+\bar{n}_2)+\gamma_3(1+\bar{n}_3)[\bar{n}_1\gamma_1+\bar{n}_2\gamma_2])\\ \nonumber
    \Tilde{\rho}_{3,3}&=\frac{1}{N}(\bar{n}_2\gamma_1 \gamma_2(1+\bar{n}_1)+\bar{n}_3 \gamma_3[\bar{n}_1\gamma_1+\bar{n}_2\gamma_2])\\ \nonumber
\end{align}
and $\Tilde{\rho}_{1,2}= \Tilde{\rho}_{2,1}=\Tilde{\rho}_{1,3}=\Tilde{\rho}_{3,1}=\Tilde{\rho}_{2,3}=\Tilde{\rho}_{3,2}=0$. In the above equations we introduced the normalization factor:
\begin{align}
    N :=& \gamma_1 \gamma_3(1+2\bar{n}_1+\bar{n}_3+3\bar{n}_1 \bar{n}_3) + \gamma_2 \gamma_3(\bar{n}_2+\bar{n}_3+3\bar{n}_2 \bar{n}_3) \notag \\ &+\gamma_1\gamma_2(1+2(\bar{n}_1+\bar{n}_2) +3 \bar{n}_1 \bar{n}_2),
\end{align}
such that $\bar{n}_i=[\exp{(\beta_i B_i)}-1]^{-1}$ models the thermal occupation of the baths.

\section{Detailed expressions for the quantum thermodynamic quantities}\label{ap:power}

Here, we give the detailed expressions for the average heat currents absorbed from reservoir auxiliary units, and average coherent power, respectively, after taking the limit $\tau \rightarrow 0$:
\begin{align}
\dot{Q}_i = \lim_{\tau \rightarrow 0} \frac{Q_i}{\tau}, ~~~~~~ \dot{W}_i = \lim_{\tau \rightarrow 0} \frac{W_i}{\tau},
\end{align}
From Eqs.~\eqref{eq:heat} and Eq.~\eqref{eq:coherentwork} for the exchanges in a single collision we obtain, for the heat currents from the reservoirs we obtain:
\begin{align}
    \Dot{Q}_{1}&= 2B_1 \gamma_1(\bar{n}_1 \Tilde{\rho}_{1,1}-(1+\bar{n}_1)\Tilde{\rho}_{2,2}),\\
    \Dot{Q}_{2}&= 2B_2 \gamma_2(\bar{n}_2 \Tilde{\rho}_{1,1}-(1+\bar{n}_2)\Tilde{\rho}_{3,3}), \\
    \Dot{Q}_{3}&= 2B_3 \gamma_3(\bar{n}_3 \Tilde{\rho}_{2,2}-(1+\bar{n}_3)\Tilde{\rho}_{3,3}), 
\end{align}
and the coherent power contributions:
\begin{align}        
    \Dot{W}_{1}&= i B_1  \sqrt{2(1+2 \bar{n}_1)\gamma_1}\lambda_1 (e^{i \phi_1}\Tilde{\rho}_{2,1}-e^{-i \phi_1}\Tilde{\rho}_{1,2} ), \\
    \Dot{W}_{2}&= i B_2  \sqrt{2(1+2 \bar{n}_2)\gamma_2}\lambda_2 (e^{i \phi_2}\Tilde{\rho}_{3,1} -e^{-i \phi_2}\Tilde{\rho}_{1,3}), \\
    \Dot{W}_{3}&= i B_3 \sqrt{2(1+2 \bar{n}_3)\gamma_3}\lambda_3 (e^{i \phi_3}\Tilde{\rho}_{3,2} -  e^{-i\phi_3}\Tilde{\rho}_{2,3}).
\end{align}
where in both sets of equations we assumed all the reservoirs are eventually injected with coherence, and steady-state conditions. The explicit expressions for the heat currents and coherent power contributions can be obtained by replacing the expression of the steady state density matrix elements in Eq.~\eqref{ness} for the general case in the above equations.

For instance, in the following we give explicit expressions in the case of coherence in the cold bath for the heat current from the cold bath $\dot{Q}_1$ and the corresponding coherent power $\dot{W}_1$. Similar expressions can be obtained for the remaining heat currents $\dot{Q}_2$ and $\dot{Q}_3$, as well as for the cases in which coherence is injected in the hot or interemediate temperature baths.
\begin{widetext}
\begin{align}
     \Dot{Q}_1 &= 4 B_1 \gamma _1 \left(-\frac{A}{C}-\frac{\left(\bar{n}_1+1\right) \left(\bar{n}_2
   \gamma _2+\gamma _2+\gamma _3+\gamma _3 \bar{n}_3+\frac{L}{M}\right)}{\gamma _2
   \left(\bar{n}_2+1\right)+\gamma _3 \left(2 \bar{n}_3+1\right)}\right); ~~~~~~
\Dot{W}_1 =\frac{X}{Y+Z+\Omega},
\end{align}
with

\begin{align}
    A &= \bar{n}_1 \{8 \gamma _1 (2 \bar{n}_1+1) [\gamma _1
   (2\bar{n}_1+1)+\gamma _2 \bar{n}_2 + \gamma _3 \bar{n}_3] [\gamma _2
   (\bar{n}_2+1) + \gamma _3 (\bar{n}_3+1)] \lambda _1^2 +4
   [B_1^2+(\gamma _1 (2 \bar{n}_1+1) \\ 
   &+\gamma _2 \bar{n}_2+\gamma _3
   \bar{n}_3){}^2] [\gamma _2 \gamma _3 (\bar{n}_2+1) \bar{n}_3+\gamma
   _1 (\bar{n}_1+1) (\gamma _2 (\bar{n}_2+1) +\gamma _3 (\bar{n}_3+1))]\}, \nonumber
\end{align}
\begin{align}
   C &=-8 \gamma _1 (2 \bar{n}_1+1)
   [\gamma _1 (2 \bar{n}_1+1)+\gamma _2 \bar{n}_2+\gamma_3 \bar{n}_3]
  [\gamma _2 (3 \bar{n}_2+2)+\gamma _3 (3 \bar{n}_3+2)] \lambda
   _1^2  -4 [-i B_1+\gamma _1 (2 \bar{n}_1+1)\\
   &+\gamma _2 \bar{n}_2+\gamma _3
  \bar{n}_3] [i B_1+\gamma _1 (2 \bar{n}_1+1)+\gamma _2 \bar{n}_2+\gamma _3
   \bar{n}_3] [\gamma _2 \gamma _3 (3 \bar{n}_3
   \bar{n}_2 +\bar{n}_2+\bar{n}_3) +\gamma_1 (\gamma _2 (3 \bar{n}_2
   \bar{n}_1+2 \bar{n}_1+2 \bar{n}_2+1) \nonumber \\
   &+\gamma _3 (3 \bar{n}_3 \bar{n}_1+2 \bar{n}_1+\bar{n}_3+1))], \nonumber
\end{align}
\begin{align}
L &=(\gamma _2 (2
   \bar{n}_2+1)+\gamma _3 (\bar{n}_3+1)) \{ 8 \gamma _1 (2
  \bar{n}_1+1) [\gamma_1 (2 \bar{n}_1+1)+\gamma_2 \bar{n}_2+\gamma_3
   \bar{n}_3] [\gamma _2 (\bar{n}_2+1) + \gamma _3
   (\bar{n}_3+1)] \lambda _1^2 \\ \nonumber
   &+4 [B_1^2+(\gamma_1(2
  \bar{n}_1+1)+\gamma_2 \bar{n}_2+\gamma _3 \bar{n}_3){}^2] [\gamma _2
   \gamma _3 (\bar{n}_2+1)\bar{n}_3 +\gamma _1 (\bar{n}_1+1) (\gamma _2
  (\bar{n}_2+1)+\gamma_3 (\bar{n}_3+1))] \},
\end{align}
\begin{align}
    M =& -8 \gamma _1
   (2 \bar{n}_1+1) [\gamma _1 (2 \bar{n}_1+1)+\gamma _2
   \bar{n}_2+\gamma _3 \bar{n}_3] [\gamma _2 (3 \bar{n}_2+2)+\gamma _3
   (3 \bar{n}_3+2)] \lambda _1^2 \\ 
   &-4 [-i B_1+\gamma _1 (2
   \bar{n}_1+1)+\gamma _2 \bar{n}_2+\gamma _3 \bar{n}_3] [i B_1+\gamma _1 (2
   \bar{n}_1+1)+\gamma _2 \bar{n}_2+\gamma _3 \bar{n}_3][\gamma _2 \gamma _3
   (3 \bar{n}_3 \bar{n}_2 \nonumber \\
   &+\bar{n}_2+\bar{n}_3)  
   +\gamma _1 (\gamma _2 (3
   \bar{n}_2 \bar{n}_1+2 \bar{n}_1+2 \bar{n}_2+1)+\gamma _3 (3 \bar{n}_3
  \bar{n}_1+2 \bar{n}_1+\bar{n}_3+1))], \nonumber
\end{align}

and 
\begin{align}
    X &=8 B_1 \gamma _1 \lambda _1^2 (2 \bar{n}_1+1) [\gamma _1 (2
   \bar{n}_1+1)+\gamma _2 \bar{n}_2+\gamma _3 \bar{n}_3] [\gamma _2 \gamma _3
   (\bar{n}_3-\bar{n}_2)+\gamma _1 (\gamma _2 (\bar{n}_2+1)+\gamma _3(\bar{n}_3+1))],
\end{align}
\begin{align}
    Y &=\gamma _2 \gamma _3 [\bar{n}_3+\bar{n}_2 (3 \bar{n}_3+1)] [(\gamma _2
   \bar{n}_2+\gamma _3 \bar{n}_3){}^2+B_1^2]+\gamma _1^2 (2 \bar{n}_1+1) [\gamma
   _2 (2 \bar{n}_1+1) (3 \gamma _3 (\bar{n}_2+(3 \bar{n}_2+1)
   \bar{n}_3)  \\
   &+\lambda _1^2 (6 \bar{n}_2+4))+2 \gamma _3 (\gamma _3 \bar{n}_3
   (\bar{n}_3+\bar{n}_1 (3 \bar{n}_3+2)+1)+\lambda _1^2 (2 \bar{n}_1+1)
   (3 \bar{n}_3+2)) \nonumber \\
   &+2 \gamma _2^2 \bar{n}_2 (2 \bar{n}_2+\bar{n}_1(3
   \bar{n}_2+2)+1)] +\gamma _1^3 (2 \bar{n}_1+1){}^2 [\gamma _2 (2
   \bar{n}_2+\bar{n}_1 (3 \bar{n}_2+2)+1)+\gamma _3 (\bar{n}_3+\bar{n}_1 (3
   \bar{n}_3+2)+1)], \nonumber
\end{align}
\begin{align}
Z &=\gamma _1 \{\gamma _2^2 \bar{n}_2 [(2 \bar{n}_1+1) (3 \gamma _3 \bar{n}_2+\lambda
   _1^2 (6 \bar{n}_2+4))+\gamma _3 (11 \bar{n}_2+\bar{n}_1 (21
   \bar{n}_2+8)+4) \bar{n}_3] +\gamma _2^3 \bar{n}_2^2 (2 \bar{n}_2  
   +\bar{n}_1 (3
   \bar{n}_2+2)+1)\}, 
\end{align}
\begin{align}
    \Omega &= \gamma _1 \{\gamma _2 [\bar{n}_1 (\bar{n}_2 (3 B_1^2+8 \gamma _3 \lambda _1^2)+8
   \gamma _3 \bar{n}_3 (\bar{n}_2 (\gamma _3+3 \lambda _1^2)+\lambda _1^2)+3 \gamma
   _3^2 (7 \bar{n}_2+2) \bar{n}_3^2+2 B_1^2)+2 \bar{n}_2 (\gamma _3 (\gamma _3
   \bar{n}_3 (5 \bar{n}_3+2) \\
   &+\lambda _1^2 (6 \bar{n}_3+2))+B_1^2)+\gamma
   _3 \bar{n}_3 (3 \gamma _3 \bar{n}_3+4 \lambda _1^2)+B_1^2] +\gamma _3 [\bar{n}_1
   (3 \bar{n}_3+2) (\gamma _3 \bar{n}_3 (\gamma _3 \bar{n}_3 +4 \lambda
   _1^2)+B_1^2)+\bar{n}_3 (\gamma_3 (\gamma_3 \bar{n}_3
   (\bar{n}_3 \nonumber \\ 
   &+1)+\lambda _1^2 (6 \bar{n}_3+4))+B_1^2)+B_1^2]\}. \nonumber
\end{align}

\end{widetext}

\section{Non-equilibrium transition points}\label{App:lambda}
When the reservoirs are thermal, we defined in Eq.~\eqref{eq:btransition} the transition points where the machine changes its regime of operation. Here we give the explicit expressions for the values of the coherence parameter $\lambda_i$ with $i = 1, 2, 3$ signaling transitions between the different combined and hybrid operation regimes appearing when coherence is present. These are obtained from the analytical expressions for the heat currents and coherent power (see Appendix~\ref{ap:power}): 
\begin{widetext}
\begin{align}
    \lambda_1^{*}&=\frac{\sqrt{(-\bar{n}_1 \bar{n}_3 + \bar{n}_2 [1 + \bar{n}_1 + \bar{n}_3]) (B_1^2 + [(1 + 2 \bar{n}_1)\gamma_1 + \bar{n}_2 \gamma_2 + \bar{n}_3 \gamma_3]^2)}}{\sqrt{2 (1 + 2 \bar{n}_1) (-\bar{n}_2 + \bar{n}_3) ([1 +2\bar{n}_1] \gamma_1 + \bar{n}_2 \gamma_2 + \bar{n}_3 \gamma_3)}},\\
    \lambda_2^{*}&= \frac{\sqrt{(-\bar{n}_1 \bar{n}_3 + \bar{n}_2 [1 + \bar{n}_1 + \bar{n}_3])(B_2^2 + [\bar{n}_1 \gamma_1 + (1 + 2 \bar{n}_2) \gamma_2 + (1 + \bar{n}_3) \gamma_3]^2)}}{\sqrt{-2 (1 + 2 \bar{n}_2) (1+\bar{n}_1 + \bar{n}_3) (\bar{n}_1 \gamma_1 + (1 + 2 \bar{n}_2) \gamma_2 + [1 + \bar{n}_3]\gamma_3)}},\\
    \lambda_3^{*}&=\frac{\sqrt{(-\bar{n}_1 \bar{n}_3 + \bar{n}_2 [1 + \bar{n}_1 + \bar{n}_3])(B_3^2 + [(1 + \bar{n}_1)\gamma_1 + (1 + \bar{n}_2) \gamma_2 + (1 + 2 \bar{n}_3) \gamma_3]^2)}}{\sqrt{2 (\bar{n}_1 - \bar{n}_2) (1 + 2\bar{n}_3) ((1 + \bar{n}_1) \gamma_1 + (1 + \bar{n}_2) \gamma_2 + (1 + 2 \bar{n}_3) \gamma_3)}}, \\
     \lambda_1^{NE}&= \frac{\sqrt{\gamma_2 \gamma_3([-\bar{n}_1 \bar{n}_3 + \bar{n}_2 (1 + \bar{n}_1 + \bar{n}_3)])(B_1^2 + [(1 + 2 \bar{n}_1) \gamma_1 + \bar{n}_2 \gamma_2 + \bar{n}_3 \gamma_3]^2)}}{\sqrt{2(-[(1 + 2 \bar{n}_1) \gamma_1 ((1 + 2 \bar{n}_1) \gamma_1 + \bar{n}_2 \gamma_2 +\bar{n}_3 \gamma_3) ((1 + \bar{n}_2) \gamma_2 + (1 + \bar{n}_3) \gamma_3)])}},\\
     \lambda_2^{NE}&=\frac{\sqrt{((-\bar{n}_1 \bar{n}_3 + \bar{n}_2 [1 + \bar{n}_1 + \bar{n}_3])\gamma_1 \gamma_3)(B_2^2 + [\bar{n}_1 \gamma_1 + (1 + 2 \bar{n}_2) \gamma_2+ (1 + \bar{n}_3) \gamma_3]^2)}}{\sqrt{2 ((1 + 2 \bar{n}_2) \gamma_2 ((1 + \bar{n}_1) \gamma_1 + \bar{n}_3 \gamma_3) (\bar{n}_1\gamma_1 + (1 + 2 \bar{n}_2) \gamma_2 + (1 + \bar{n}_3) \gamma_3))}}, \\ 
     \lambda_3^{NE}&=\frac{\sqrt{((-\bar{n}_1 \bar{n}_3 + \bar{n}_2 [1 + \bar{n}_1 + \bar{n}_3])\gamma_1 \gamma_2)(B_3^2 + [(1 + \bar{n}_1) \gamma_1 + (1 + \bar{n}_2) \gamma_2 + (1 + 2 \bar{n}_3) \gamma_3]^2)}}{\sqrt{2(-[(1 + 2 \bar{n}_3) (\bar{n}_1 \gamma_1 + \bar{n}_2 \gamma_2) \gamma_3 ((1 +\bar{n}_1) \gamma_1 + (1 + \bar{n}_2) \gamma_2 + (1 + 2 \bar{n}_3) \gamma_3)])}}, 
\end{align}
where $\lambda_i^{*}$ and $\lambda_i^{NE}$ are obtained by solving the equations for the heat fluxes $\Dot{Q}_{j \neq i}=0$ and $\Dot{Q}_i=0$ respectively, when coherence is present in the $i$th reservoir. 

The above curves are generic and provide the transition lines between regimes of operation of the coherent machine, as illustrated in Fig.~\ref{fig:cohbaths}. There $\lambda_1^{NE}$ separates regime III from regime VII as seen in Fig.~\ref{fig:cohbaths}a), while the line $\lambda_1^{*}$ corresponds the transition from regimes III-VII to regime IV. In Fig.~\ref{fig:cohbaths}b), when coherence is in the intermediate bath, $\lambda_2^{NE}$ separates regime IV from regimes III-VIII. On the other hand, the curve $\lambda_2^{*}$ separates regime VIII from regime III. In Fig.~\ref{fig:cohbaths}c), the boundary between regimes IV and regime V is given by $\lambda_3^{*}$. Moreover, the curve in $\lambda_3^{NE}$ separates regimes III and V. The transition curve between regimes IV and III occurs when $\lambda_3^{*} \approx \lambda_3^{NE}$.
\end{widetext}

\section{Coherence with equal bath temperatures}\label{Ap:sametemp}

\begin{widetext}
\begin{figure*}[htp]
\begin{center}
\subfigure[]{\includegraphics[height=4.5cm]{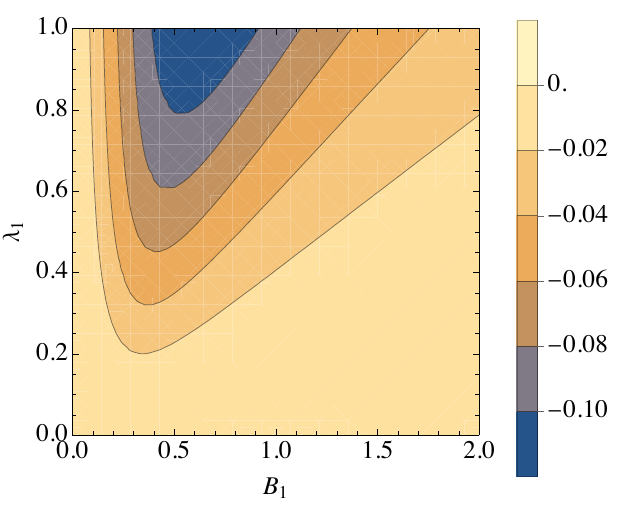}}
\subfigure[]{\includegraphics[height=4.2cm]{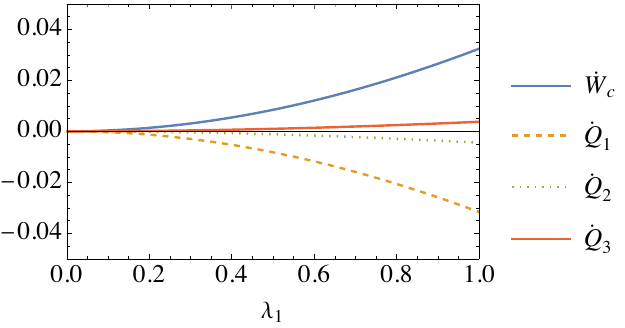}}
\caption{(a) Contourplot of $\Dot{Q}_1$ in units of $T_1 \gamma_1$ as a function of $B_1$ and $\lambda_1$ when $T_1=T_2=T_3=6$. (b) Plot of all thermodynamic quantities in terms of $\lambda_1$ for a fixed value of $B_1=2$, the rest of the parameters are: $\gamma_1= 8.7 \times 10^{-3}$, $\gamma_2 = 5.7 \times 10^{-3}$, and $\gamma_3 = 7.5 \times 10^{-3}$.}
\label{fig:coldbathcoherence1}
\end{center}
\end{figure*}
\end{widetext}

The presence of coherence within the environmental units of the baths can lead to heat currents through the system in the NESS, even when there is no temperature gradient between the three terminals. This point is illustrated in Fig.~\ref{fig:coldbathcoherence1} obtained by setting equal temperatures in the three baths.  In Fig.~\ref{fig:coldbathcoherence1}a), we plot $\Dot{Q}_1$ as a function of $B_1$ and $\lambda_1$ for the case of coherence in the bath $1$ (notice that this bath can no longer be considered the ``cold bath" as before)  where $\Dot{Q}_1$  takes non-zero values. Furthermore, Fig.~\ref{fig:coldbathcoherence1}b) displays all energetic currents for a constant value of $B_1$. As can be observed from the signs of the currents, in this situation as $\lambda_1$ increases the input coherent power from environmental coherence leads to a heat current from reservoir $3$ to reservoirs $2$ and $1$, without any temperature gradient in the baths. 

 \bibliography{biblio1}



\end{document}